# A Trigger Sequence in a Leucine Zipper Aids its Dimerization; Simulation Results

Robert I. Cukier*

Department of Chemistry

Michigan State University, East Lansing, MI 48824-1322

**Keywords**: Leucine Zipper; Molecular Dynamics; Replica Exchange Method; trigger sequence

*Corresponding author:

Professor R. I. Cukier

Department of Chemistry

Michigan State University

E. Lansing, MI 48224-1322

cukier@chemistry.msu.edu

Phone: 517-353-1170

Fax: 517-353-1793

**Abstract**

Leucine zippers are alpha helical monomers dimerized to a coiled coil structure. Various scenarios for dimerization span interaction of unstructured monomers that form alpha helices in the process of dimerization to preformed alpha helical monomers dimerizing. Another suggested possibility, a "trigger sequence" hypothesis (M. O. Steinmetz et al., *Proc. Natl. Acad. Sci. U. S. A.* 2007, 104, 7062-7067), is that a C terminal (CT) part of each monomer is a trigger sequence that is dimerized and, subsequently, the remainder N terminal (NT) part of each monomer (that is partially ordered) zips together to form the coiled coil. In this work, methods are developed to computationally explore the trigger sequence hypothesis based on an extension of a previously introduced (R. I. Cukier, J. Chem. Phys. 2011, 134, 045104) Hamiltonian Temperature Replica Exchange Method Mean Square (HTREM_MS) procedure, which scales the Hamiltonian in both potential and kinetic energies, to enhance sampling generically, along with additional restraints to enhance sampling in desired regions of conformational space. The method is applied to a 31-residue truncation of the 33-residue leucine zipper (GCN4-p1) of the yeast transcriptional activator GCN4. Using a variety of HTREM_MS simulations, we find that the NT of one monomer of the dimer becomes disordered when the NT parts of the monomers are separated. In contrast, when the CT parts of the monomers are separated, both monomers remain alpha helical. These simulations suggest that the CT does act as a trigger sequence and are consistent with a disordered NT. We also investigate whether the disordered NT can be induced to re-form the dimerized leucine zipper. It does, but with some lack of recovery of alpha helical hydrogen bonding structure in the NT.

## 1. INTRODUCTION

Leucine zippers (LZs)[1-9] are an important class of regulatory proteins that are integral to e.g. eukaryotic transcription factors.[10] They consist of dimerized (or oligomerized) alpha helices that form coiled-coil motifs. Crick's[11] "knobs-into-holes" model for coiled-coil dimerization emphasized the role of packing, but electrostatic interactions also are important to dimer stability.[5, 9, 12-15] Folding pathways and stabilities of GCN4, a yeast transcriptional activator leucine zipper, and its mutants, have been investigated using calorimetry, circular dichroism, kinetics, hydrogen exchange, and NMR.[14-20] Numerous computational studies using lattice models[21-24], coarse grained models[25], implicit solvent[26-27], and explicit solvent[28-30] simulations of monomer folding and dimerization of GCN4-p1 and other leucine zippers have been carried out.

Leucine zipper dimerization is a specific instance of protein-protein interaction that is viewed as formation of an encounter complex by diffusion (with, potentially, an electrostatic steering component)[31] followed by the overcoming of a free energy barrier to form the bound, stable state. For proteins, the individual proteins (monomers) must undergo complex rearrangements in the process of forming the stable complex (dimer).[32] One scenario is "lock-and-key" in which preformed monomers only need to fit together. A different scenario involves an encounter complex of essentially random coil monomers that then extensively reorganize (fold) to dimerize. An intermediate scenario related to intrinsically disordered proteins[33-34] would be for a part of one monomer to be "disordered" when separated from its partner monomer and becomes ordered upon binding with its partner monomer.

These considerations arise in the context of LZ dimerization. The GCN4-p1 leucine zipper[35] that is the dimerization domain of the DNA-binding protein GCN4[36], is an example of an alpha helical, parallel, double-stranded, coiled-coil. Each monomer consists of 31 residues that can be divided into two domains: an N-terminal (NT), residues 1-15, and a C-terminal (CT), residues 16-31 domain. The sequence is:

{R[M(*a*)KQL(*d*)EDK][V(*a*)EEL(*d*)LSK]}{[N(*a*)YHL(*d*) ENE][V(*a*)ARL(*d*)KKL]V(*a*)G}.

The "a" and "d" indicate the hydrophobic residues, the leucine repeats are in [], and the NT and CT domains are indicated by the {}. It was suggested[15] that the CT domain of the LZ is a trigger sequence that, when dimerized, would help "zip up" the NT monomers to form the GCN4-p1 dimer, and that a trigger sequence is a general phenomenon.[37] The electrostatic and hydrogen bonding properties of a 16 residue peptide corresponding to residues 16-31 of GCN4-p1 were investigated to ascertain properties of this trigger sequence. The dimerized CT[15] may drive coiled-coil formation by limiting the available configuration space prior to complete dimerization.[30] Study of a hybrid sequence composed in part of GCN4 and cortexillin[38] supported the importance of a stabilizing (versus a specific trigger) sequence to coiled-coil dimerization. Whether the properties of the trigger sequence are quite specific or not is not clear but the trigger sequence concept seems quite robust,[39-41] leading to the conclusion that GCN4-p1 folds by diffusional encounter of preformed CT helices that nucleate the dimerization of more disordered NT residues.[41]

In this work, we computationally investigate aspects of the above scenario. From the perspective of Molecular Dynamics (MD) simulations of leucine zippers both melting (ordered dimer to disordered monomers) and interface formation (disordered monomers to ordered dimer) are challenging due to the variety of barriers that arise from the high-dimensional potential energy surface and the large configuration space that must be traversed. Thus, it is useful to apply some form of accelerated MD. To this end, we previously introduced a variant of the Hamiltonian Replica Exchange Method (HREM)[42-43], designated as HTREM_MS[44] that: 1) scales selected terms in the Hamiltonian in both potential and kinetic energies, the combination defining HTREM, and 2) introduces specific distance restraints (MS designates mean square restraints). HTREM limits the number of scaled degrees of freedom, resulting in a smaller number of systems required relative to temperature REM[45-48] where all degrees of freedom are scaled. Recent work[49] has shown that use of a similar strategy[50] can greatly enhance sampling in the degrees of freedom orthogonal to those of the applied restraints. In a previous LZ HTREM_MS simulation[44] that showed that the dimer interface could be reformed from separated, alpha-helical monomers, restraints between specific monomer-monomer distances served to encourage the monomers to sample more dimer-like conformers. Here, using a variety of HTREM_MS, we investigate whether the LZ NT is stable when those parts of the monomers are separated, whether the CT is stable when those parts of the monomers are separated, and whether a melted NT can be induced to re-form the dimer. HRRE These simulations suggest that: 1) the CT does act as a trigger sequence and permits part of the NT to melt, 2) separating the CT does not lead to significant melting of either the NT or CT domains and 3) starting from a melted NT, the dimer does reform with, however, some lack of alpha helical hydrogen bond recovery in the NT.

The plan of the rest of this paper follows. Section 2 provides a brief summary of the HTREM_MS method as appropriate to the current study, along with some simulation details, and an overview of the simulations carried out. The hydrogen bond and salt bridge criteria are defined. Section 3 describes the details of the simulations of dimer stability, NT separation, CT separation and restoration of the dimer from the NT melted leucine zipper. A discussion of these results and conclusions that can be drawn from the study are presented in Section 4.

## 2. METHODS

**2.1. HTREM_MS.** In a replica exchange method (REM) the basic strategy is to create a set of MD systems characterized by unique Hamiltonians. These are run in parallel and periodically information exchange between pairs of systems is attempted and accepted according to a Monte Carlo rule. If successful, the method will provide Boltzmann sampling in all the systems and the sampling will be accelerated relative to conventional MD. If one of the systems is simulated at a temperature of interest and is not restrained, it will characterize the Boltzmann sampling of the physical system.

The HTREM_MS method is a Hamiltonian TREM (HTREM) with the addition of restraint potentials. The HTREM Hamiltonians[43] are $H_\lambda(\mathbf{X},\mathbf{P}) = K_\lambda(\mathbf{P}) + V_\lambda(\mathbf{X})$ where $K_\lambda(\mathbf{P})$ is the kinetic energy and $V_\lambda(\mathbf{X})$ is the potential function for a system parameterized with λ, with phase space coordinates $\mathbf{X}, \mathbf{P} \equiv \Gamma$. For HTREM_MS, $H_{\lambda,\mu} = H_\lambda + V_\mu^r$ with the $V_\mu^r$ a set of restraint potentials parameterized by $\mu$ that are designed to enforce e.g. desired monomer-monomer residue separations.

Between exchange attempts MD is run for each system characterized by its $H_{\lambda,\mu}$. When system interchanges are to be attempted, detailed balance is enforced. In our HTREM scheme the HTREM potential is written as

$$V_\lambda(\mathbf{X}) = \lambda V_{PP}(\mathbf{x}_{PP}) + \sqrt{\lambda} V_{PS}(\mathbf{x}_{PS}) + V_{SS}(\mathbf{x}_{SS}). \tag{2.1}$$

with subscript "P" for protein and subscript "S" for solvent. The protein-protein interactions are scaled with $\lambda$ and the protein solvent with $\sqrt{\lambda}$. An effective temperature $T_{eff}$ can be defined[43] from the number of scaled degrees of freedom, for a given value of $\lambda$. In contrast with temperature REM, fewer systems are needed as fewer degrees of freedom are scaled. The form of the potential stems from the Ewald method requirement of overall charge neutrality. Use of the Metropolis rule to govern the acceptance probability α between any two systems

$$\alpha(\Gamma, \Gamma' \to \Gamma', \Gamma) = \min\left(1, e^{-\Delta(\Gamma, \Gamma' \to \Gamma', \Gamma)}\right) \tag{2.2}$$

where

$$\Delta(\Gamma, \Gamma' \to \Gamma', \Gamma) = \beta\left[\left(H_{\lambda,\mu}(\Gamma') + H_{\lambda',\mu'}(\Gamma)\right) - \left(H_{\lambda,\mu}(\Gamma) + H_{\lambda',\mu'}(\Gamma')\right)\right] \tag{2.3}$$

guarantees that Boltzmann equilibrium results in the extended ensemble of the product of all the systems' ensembles, for a sufficiently long trajectory.

The restraint potential used here is a sum over half-harmonic pair potentials between selected atoms in each monomer:

$$V_r^{\pm}(\mu) = \frac{1}{2}\sum_j k_j \left(r_j - r_j^t\right)^2 \Theta_{\pm}\left(r_j - r_j^t\right) \tag{2.4}$$

with $r_j^t$ the $j$th target distance and $\Theta_+(z)$ ($\Theta_-(z) = 1 - \Theta_+(z)$) a step function. Each term in the sum is half-harmonic in the instantaneous distance between an atom pair relative to the target distance, with the non-zero force constant branch of $V_r^+$ ($V_r^-$) corresponding to distances greater than (smaller than) the target distances, with the idea to encourage bringing the monomers together (encourage separating the monomers).

Note that, as in any REM-based simulation, the exchange attempts can be thought of as exchanges of configuration between systems with different parameters, or as exchanges of parameters between indexed configurations (corresponding to MD running on different processors). To the extent that large moves in configuration space occur when systems exchange their parameters ($\lambda$ and µ) the sampling rate should be enhanced. In our implementation, the systems are run independently on different nodes of a Linux cluster computer and, when exchanges attempts are successful, the $\lambda, \mu$ values are passed using the Message Passing Interface technique implemented as MPICH, permitting frequent (here, every 40 MD steps) attempts at exchange without significant loss of computational efficiency.

**2.2 MD protocol for the Leucine Zipper.** The CUKMODY[51] protein molecular dynamics code that uses the GROMOS96[52] force field was used for all simulations; the conditions are detailed elsewhere.[43] The starting GCN4-p1 leucine zipper dimer configuration was obtained from its x-ray structure (PDB accession code 2ZTA).[35] The crystal structure construct has 33 residues (in each monomer). The last two were not resolved and considered to be in a random-coil configuration. Thus we simulated residues 1-31 that were resolved. The first two (NT) residues have higher crystallographic B values and the phi/psi angles for residues 3-30 are consonant with alpha helical structure. Thus, the first two NT residues are not well ordered in the crystal structure.  After that, the phi/psi dihedral values are characteristic of alpha helices. Our simulations are in excellent agreement with these results. The simulations were carried out in a box with 59.1851 Å sides with SPC waters. Table 1 provides an overview of the simulations that were carried out. For the HTREM and HTREM_MS simulations eight windows were used. Window 0 always corresponds to the ambient temperature, T=$T_{eff}$=275, to match the temperature of the experiments.[15, 17]

The preparation phase equilibrated the dimer. To probe the stability of the dimer and assess the extent of alpha helical hydrogen bonding, simulations with no restraints at a series of effective temperatures were run. That established baseline hydrogen bonding percentages for the monomers. The NT separation simulation incorporates half harmonic restraints on some of the NT residues, with the non-zero forces on the close-in distances, to see if the NT helical structure does "melt", which it does. The CT separation analogous simulation uses half harmonic restraints on some CT residues. Finally, the separated, melted NT with intact dimerized CT was simulated with, now, non-zero-force exterior half harmonic restraints on NT monomer-monomer residue pairs to see if the dimer is restored.

**2.3 Definitions of Alpha Helical Hydrogen Bonds, dimer separation, and Salt Bridges**

Analyzer 2.0[53] was used for the hydrogen bond analysis. The alpha helical hydrogen bonds between the nitrogen of the residue *n*+4 N-H and the carbonyl of the residue *n* C-O are defined by backbone nitrogen to carbonyl oxygen distance less than 3 Å and corresponding OHN angle between 0 and 30°. The distance was picked from the crystal structure distances[35] and is clearly a very conservative definition of a hydrogen bond. To monitor the stability of the dimer character, the nine (5 *a–a'* and 4 *d–d'*) monomer–monomer CA distances of the LZ (see the sequence in the Introduction) were evaluated. Salt bridges were identified for: Arg-Glu and Arg-Asp by their CZ-CD and CZ-CG distances, respectively. For the positively charged residues, the Arg CZ atom (last carbon atom of the side-chain) and the Lys NZ atom (side chain N atom) were used. For the negatively charged residues, the Asp CD (carbonyl carbon) and Glu CG (carbonyl carbon) were used. These choices lead to a salt bridge cutoff of 5 Å, for all the combinations used, based on the geometry of salt bridges with the indicated atoms. As the salt bridge fractions are only used to compare their presence in two different simulations, within reason, the conclusions reached should not be influenced by the precise definition or choice of cutoff used.

**2.4 Error Analysis**. The statistical quality of the monomer alpha helical hydrogen bonding, nine distances between the monomers and the salt bridge trajectory data for all the simulations were assessed using block averaging.[54] This method divides a property trajectory into blocks and evaluates the standard error of the mean as a function of block size to obtain a plateau value. A plateau value indicates that on some sampling interval the data are uncorrelated samples and provide accurate estimates of the standard error. In our REM-based simulations, as shown in the SI, uncorrelated samples are obtained on a time scale of less than 100 ps.

## 3. RESULTS

### 3.0 Summary of simulations.

A summary of the sequence of simulations carried out and their main consequences is presented in Table 1.

**Table 1. Summary of simulations**

| Simulation | Purpose | Method | Result |
|---|---|---|---|
| Equilibration | Equilibration of dimer | HTREM | Equilibration |
| No restraints | dimer/Helix stability | HTREM | NT alpha helix HBs ~50% |
| NT separation | Separate NT of dimer | HTREM with half harmonic interior restraints on NT | NT separates and one helix melts |
| CT separation | Separate CT of dimer | HTREM with half harmonic interior restraints on CT | CT does not melt |
| Bring Together | Restore dimer | HTREM with half harmonic exterior restraints on NT | Dimer restored and some HBs |

**3.1. Preparation of an equilibrated dimer.** The equilibration of the dimer was carried out from the crystal structure with the monomers further separated by 1 Å and one monomer rotated by 15 degrees around its long axis with respect to the other monomer. This starting structure was then simulated with parameters given in Table 2. The effective temperatures are all close to T=275 K, which corresponds to $\lambda$=1.000 for window 0. (A $\lambda$ value of 0.95 corresponds to an effective temperature of 292 K). The reason for this equilibration method was to provide a starting set of structures that are not too prejudiced to the crystal structure. In the course of this simulation, the inter-monomer distances (as given in the Figure 1 legend) are regained on average, with (small) fluctuations in these distances.

**Table 2. Equilibration simulation**

| Simulation (time) | Purpose | Result | Restraints/FC | Window T | Window exchange |
|---|---|---|---|---|---|
| Preparation (2X8 ns) | Equilibration of dimer | Dimer Reformed | Leu5, Val9, Asn16, V23 FC=1.5 kcal/mol | W_0  1.000<br>W_1  0.999<br>W_2  0.988<br>W_3  0.973<br>W_4  0.968<br>W_5  0.952<br>W_6  0.947<br>W_7  0.931 | 0<--->1 0.88<br>1<--->2 0.92<br>2<--->3 0.83<br>3<--->4 0.89<br>4<--->5 0.84<br>5<--->6 0.88<br>6<--->7 0.85 |

Figure 1 displays the crystal structure on the left and an endpoint of the equilibration simulation, with monomer 1 (2) on the left (right) and the NT oriented upward.

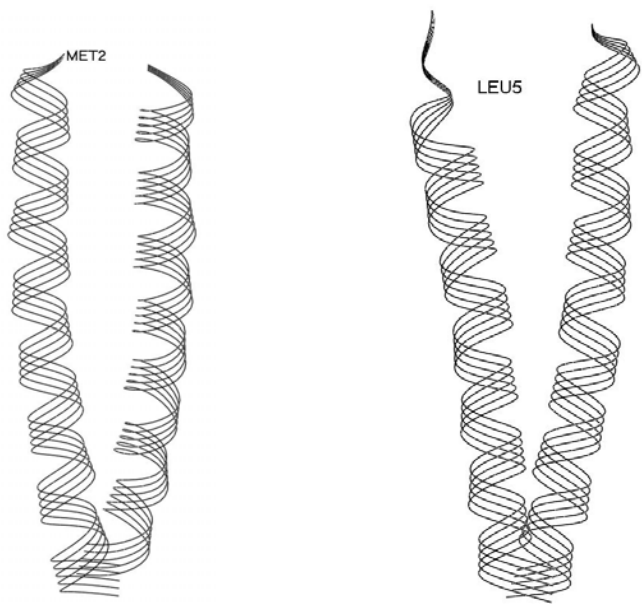

Figure 1. (Left panel) The LZ crystal structure showing the helical structure of each monomer in the coiled-coil. Met2 of monomer is indicated to show the orientation of the two monomers that is maintained in the remaining figures. (Right panel) End point of MD equilibration. The NT

residues Arg1-Leu5 are somewhat disordered here. The nine (five *a–a'* and four *d–d'*) intermonomer (alpha carbon–alpha carbon) distances (in Å) from the crystal structure[35] and their simulation averages and standard errors, given in parenthesis, are, respectively, 6.31 (7.30 ± 0.011), 6.34 (6.58 ± 0.015), 5.47 (6.06 ± 0.012), 6.11 (6.21 ± 0.011), 5.60 (6.00 ± 0.025), 6.20 (6.15 ± 0.010), 5.87 (5.93 ± 0.011), 6.43 (6.12 ± 0.010), and 6.19 (6.16 ± 0.022).

The endpoint of the equilibration simulation shows some disorder in the first few NT residues but essentially reproduces the crystal structure in terms of stable helical monomers and a bound dimer that fluctuates around the crystal structure CA-CA' distances listed in Figure 1.

**3.2 The no restraint simulation leads to a stable bound dimer.** Following the equilibration simulation, an unrestrained simulation was carried out to obtain the properties of the dimer. Here, as detailed in Table 3, a large range of temperatures was used to enhance configurational sampling. The window 7 $\lambda$ value of 0.61 corresponds to $T_{eff}$=450 K. The exchange probabilities are large indicating that many higher T configurations are feeding back to the T=275 system and result in excellent enhancement of sampling.

To check on the dimer equilibration of this no restraint simulation the window 0 ($T_{eff}$=275) system trajectory was split into two equal parts and, as in the equilibration stage, the nine dimer distances were evaluated. The nine (five *a–a'* and four *d–d'*) intermonomer (alpha carbon–alpha carbon) distances (in Å) from the crystal structure and their simulation averages and standard errors, given in parenthesis, are, respectively, 6.31 (6.56 ± 0.015), 6.34 (6.26 ± 0.007), 5.47 (5.56 ± 0.005), 6.11 (6.06 ± 0.004), 5.60 (5.58 ± 0.005), 6.20 (6.02 ± 0.004), 5.87 (5.40 ± 0.005), 6.43 (6.03 ± 0.004), and 6.12 (6.18 ± 0.011) for the first half and 6.31 (6.86 ± 0.033), 6.34 (6.15 ± 0.006), 5.47 (5.55 ± 0.005), 6.11 (6.09 ± 0.004), 5.60 (5.60 ± 0.005), 6.20 (6.01 ± 0.004), 5.87 (5.39 ± 0.005), 6.43 (6.02 ± 0.004), and 6.19 (6.18 ± 0.012) for the second

half of the trajectory. Thus, there is good agreement with the x-ray derived data, and consistency between the two half trajectories..

**Table 3. No restraint simulations**

| Simulation (time) | Purpose | Result | Window T | Window exchange |
|---|---|---|---|---|
| No restraint (8X8 ns) | dimer/helix stability | NT Alpha helix HBs ~50%. Equilibrated dimer. | W_0  1.00<br>W_1  0.94<br>W_2  0.89<br>W_3  0.83<br>W_4  0.78<br>W_5  0.72<br>W_6  0.67<br>W_7  0.61 | 0<--->1 0.53<br>1<--->2 0.51<br>2<--->3 0.52<br>3<--->4 0.55<br>4<--->5 0.57<br>5<--->6 0.57<br>6<--->7 0.59 |

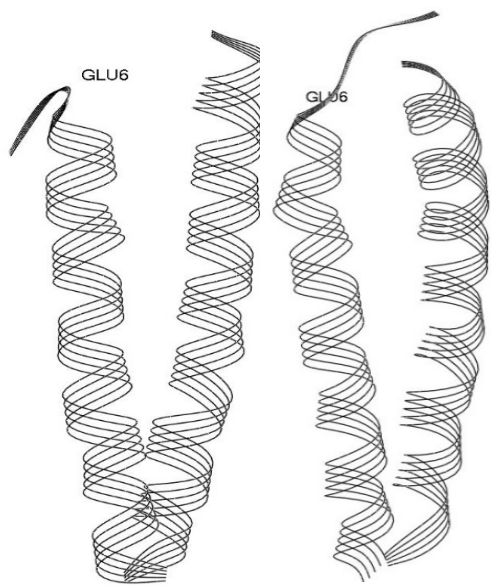

Figure 2. Two snapshots from the no restraint simulation of the dimer show that the NT of M1 is somewhat labile. The figure above is from W_0 corresponding to the ambient T.

The snapshots in Figure 2 show that there can be loss of alpha helical hydrogen bonding concentrated at the NT end. However, these configurations are not typical. Examination of a number of snapshots show that the helical structure is actually quite well maintained. A quantitative measure can be obtained from the alpha helical hydrogen bonding pattern of the NT residues. The fractions of trajectory time that there are alpha helical hydrogen bonds in monomer 1 are shown in Figure 3. Excluding the NT terminal hydrogen bond between residues 1 and 5, and the CT terminal one between residues 27 and 31, the alpha helical hydrogen bonding is present roughly one-third to two-thirds of the time. A similar pattern of hydrogen bonding presence for monomer 2 is also found (see SI Table S1). Thus, the bound leucine zipper dimer is a stable dimer of alpha-helical monomers with fluctuations of these hydrogen bonds that are essentially the same over each monomer. Note that if the strict hydrogen bond definition of Section 2.3 were loosened from the heavy atom 3.0 Å distance and OHN angle between 0 and 30°, to heavy atom 3.5 Å distance, higher probabilities, on average, ~90%, are obtained, as shown in SI Figure S1. The Steinmetz et al.[15] experiment carried out on the CT (residues 16-31) show about 50% HB based on $^{13}$CA chemical shift NMR and far-UV CD methods. The experiments of Dragan et al.[17] on the leucine zipper use CD to estimate an average of 90% alpha helical HB. Thus, the simulation does a good job of capturing both the alpha-helical character of the LZ monomers as well as their dimerization, as monitored by the nine distances, given before Table 3

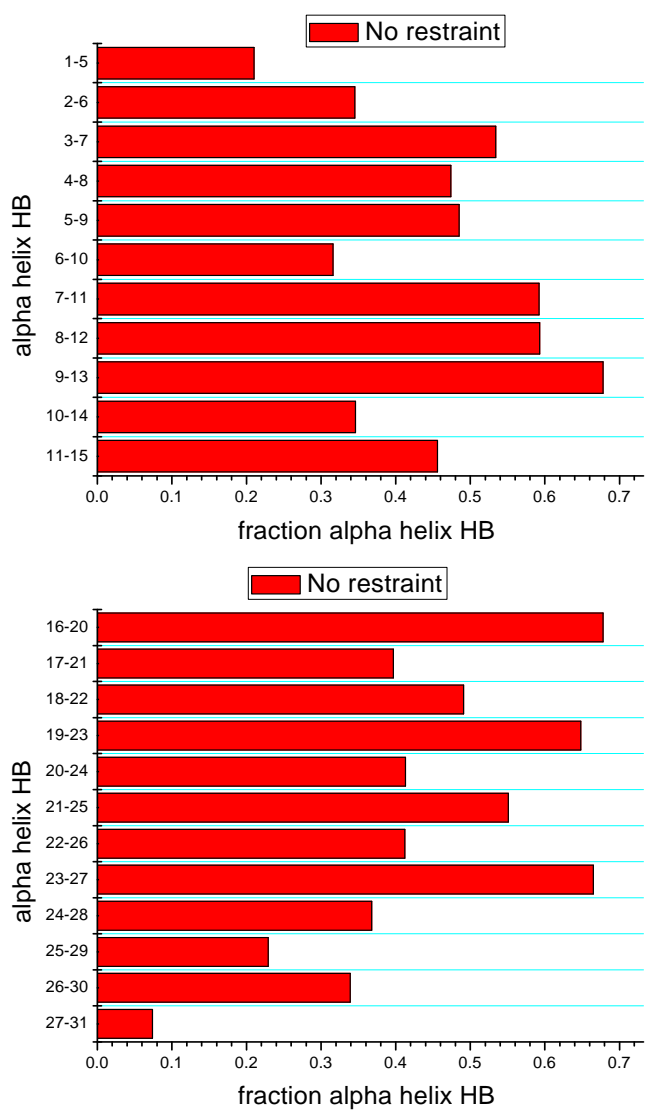

Figure 3. Fractional occurrence over the no restraint dimer trajectory (Table 3) of alpha helical hydrogen bonds between the indicated residues for the NT (top panel) and CT (bottom panel) for monomer 1 for window 0. A similar pattern is found for monomer 2 (SI Table S1). The standard errors as a function of block size are shown in the SI Figure S2. Splitting the trajectory into two equal segments and evaluating the ratios of the hydrogen bond fractions shows consistency of the above results (SI Figure S3).

**3.3 N terminus separation melts part of the monomer 1 N terminus.** With a stable, well-equilibrated dimer, restraints were added to test the CT trigger hypothesis that if the NT parts of the monomers were separated, then the alpha-helical structure of the NT might be disrupted while essentially maintaining the CT dimer structure. To this end, half harmonic restraints acting on the interior parts of selected NT inter-dimer distances were applied to all the windows. In particular, monomer-monomer' CA atom distance restraints between Leu5-Leu5', Val9-Val9', Leu12-Leu12' and Asn16-Asn16' that serve to produce additional separations of, on average, respectively, 6, 5, 4, and 3 Å were used. As shown in Table 4, for all window pairs, the exchange probabilities are excellent. The indicated residue pairs, force constants and additional distances were picked to push the NT monomers progressively further apart as the NT terminus is approached. This combination of restraints and force constants was successful in that the average distances of those restrained residues did match more-or-less their desired separations.

**Table 4. N terminus separation simulations** [a]

| Simulation (time) | Purpose | Result | Window T | Window exchange |
|---|---|---|---|---|
| Separation (4X8 ns) | Separate NT of dimer | NT helix melts partially | W_0 1.000<br>W_1 0.990<br>W_2 0.988<br>W_3 0.973<br>W_4 0.968<br>W_5 0.952<br>W_6 0.947<br>W_7 0.931 | 0<--->1 0.86<br>1<--->2 0.91<br>2<--->3 0.79<br>3<--->4 0.88<br>4<--->5 0.78<br>5<--->6 0.88<br>6<--->7 0.84 |

[a] Half harmonic interior restraints on all the windows (with FC=2.5 kcal/mol/ Å$^2$) between Leu5-Leu5', Val9-Val9', Leu12-Leu12' and Asn16-Asn16' that on average will produce additional separations of, respectively, 6, 5, 4, and 3 Å

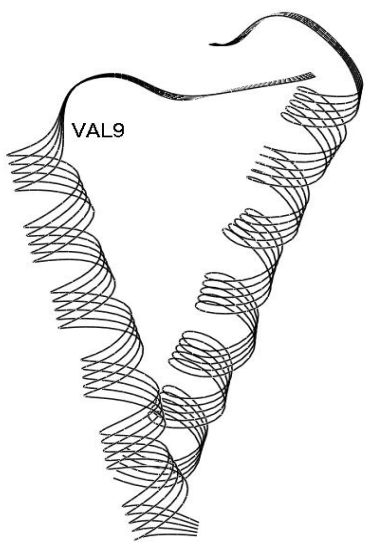

Figure 4. The separated dimers accomplished with the restraints indicated in Table 4. When the NT parts of the monomers are separated, the NT sequence melts to a certain extent. In this snapshot, the alpha-helical character of monomer 1 is lost from Val9 to the N terminus. There is also some loss of helical character in the N terminus of monomer 2.

A representative snapshot of the melted NT LZ is shown in Figure 4. The CT and parts of the NT are well maintained as alpha helices while there is evident melting in the NT spanning, in this snapshot the monomer 1 Val9 to the NT terminus. There is also some melting in monomer 2. Figure 5 displays the alpha helical hydrogen bonding pattern for monomer 1. The alpha helical character of residues 2 through 8 is lost, and this part of monomer 1 certainly can be considered as having melted. The remaining hydrogen bonds are similar in occupation to those shown in Figure 3 for the no restraint simulation. Examination of the monomer 2 alpha helix hydrogen bonds corresponding, in monomer 1, to the Met2-Glu6, Lys3-Asp7, Gln4-Lys8 hydrogen bonds show that their fractions are 0.37, 0.38 and 0.22, respectively (see SI Table S1), somewhat less than the fractional occupations for monomer 1 in the no restraint simulation (see Figure 3).

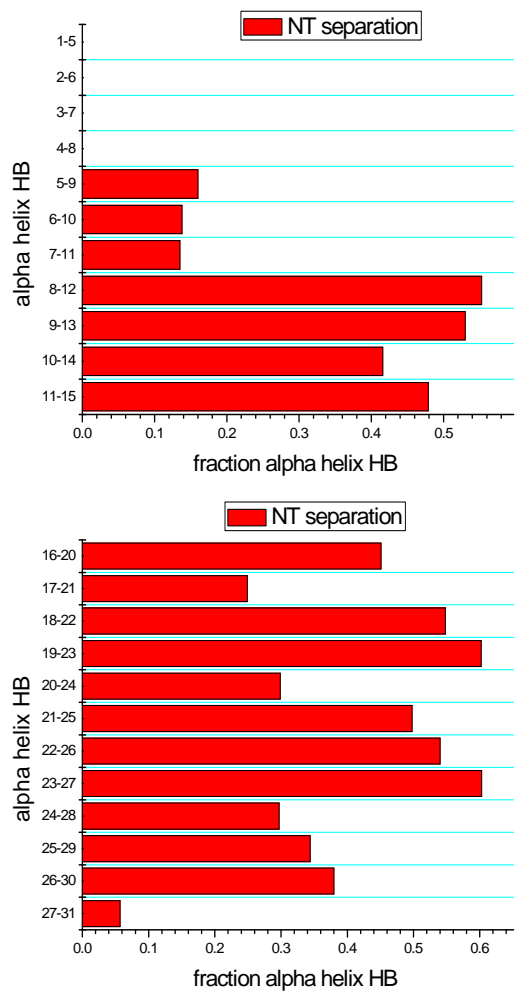

Figure 5. NT separation trajectory (see Table 4). Fractional occurrence over the trajectory of alpha helical hydrogen bonds between the indicated residues for the NT (top panel) and CT (bottom panel) for window 0. The standard errors as a function of block size are shown in the SI Figure S4. Splitting the trajectory into two equal segments and evaluating the ratios of the hydrogen bond fractions shows consistency of the above results (SI Figure S5).

**3.4 C terminus separation does not melt the C terminus, or the N terminus.** As a control simulation, a similar separation methodology as used on the NT was applied to the CT. First a run with the same window effective temperatures $T_{eff}$ as for the NT simulation of Section 3.3 was carried out. These results are summarized in the SI, Figure S6. Then, as a more severe interrogation of the stability we first used a larger $T_{eff}$ range and then collected the data over the same $T_{eff}$ as for the NT simulation.

**Table 5. C terminus separation simulations** [a]

| Simulation (time) | Purpose | Result | Window T | Window exchange |
|---|---|---|---|---|
| Separation (4X8 ns) at large T range | Separate CT of dimer | CT helix stable | W_0  1.00<br>W_1  0.94<br>W_2  0.89<br>W_3  0.83<br>W_4  0.70<br>W_5  0.72<br>W_6  0.67<br>W_7  0.61 | 0<--->1 0.661<br>1<--->2 0.649<br>2<--->3 0.579<br>3<--->4 0.665<br>4<--->5 0.569<br>5<--->6 0.671<br>6<--->7 0.587 |
| Separation (4X8 ns) at small T range | | | W_0  1.000<br>W_1  0.990<br>W_2  0.988<br>W_3  0.973<br>W_4  0.968<br>W_5  0.952<br>W_6  0.947<br>W_7  0.931 | 0<--->1 0.878<br>1<--->2 0.936<br>2<--->3 0.819<br>3<--->4 0.873<br>4<--->5 0.830<br>5<--->6 0.883<br>6<--->7 0.831 |

[a] Half harmonic interior restraints on all the windows (with FC=2.5 kcal/mol/ Å$^2$) between Asn16- Asn16', Leu19- Leu19', Val23- Val23' and Leu26-Leu26'

Table 5 summarizes these simulations, which used half harmonic interior restraints on four residue pairs were applied to all the windows with the same force constant as for the NT separation simulations. They were first run at series of effective temperatures with the highest (W_7) corresponding to $T_{eff}$=450 K. This was followed by a more restrained range of temperatures to collect more data. In all simulations the exchange fractions are very high, indicating that, even for the first set of simulations, there should be excellent HTREM sampling enhancement. The result of this simulation is that the CT terminus does not melt, in agreement with the suggestion that CT acts as a trigger sequence.

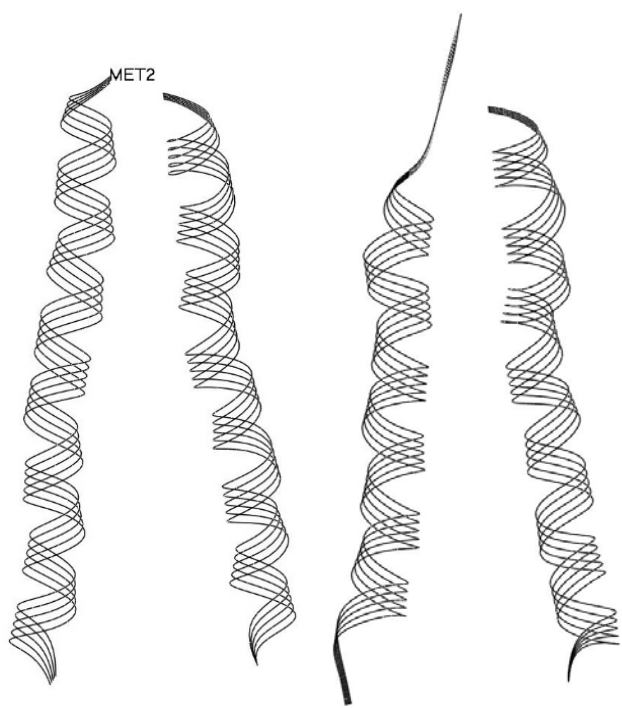

Figure 6. Two snapshots from the separation of the dimers applied to the CT of the LZ. In this case, the monomers of the separated dimer do not extensively lose their alpha helical character. The NT of monomer 1 does show some fluctuations in its conformation.

Two snapshots from this simulation are displayed in Figure 6, where it is evident that the alpha-helical structure of the monomers is well maintained, with some melting in the monomer 1 NT. The end residues of CT can show some non-helical behavior in accord with that found for the bound dimer simulation. The alpha helical fractions are plotted In Figure 7.

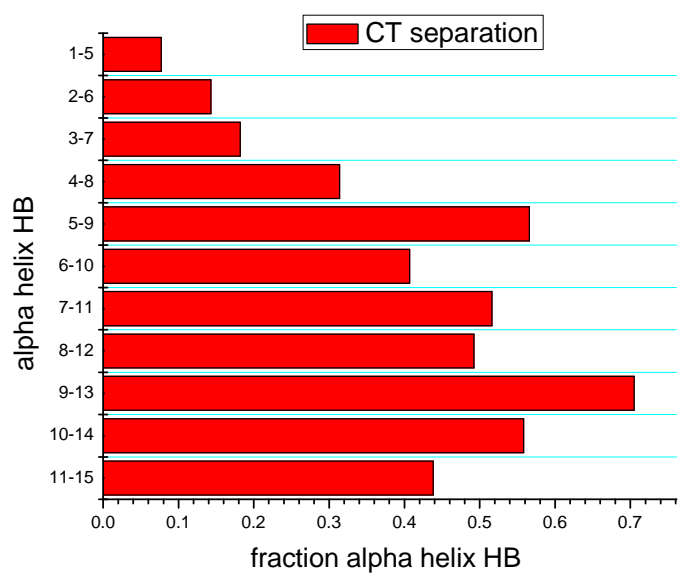

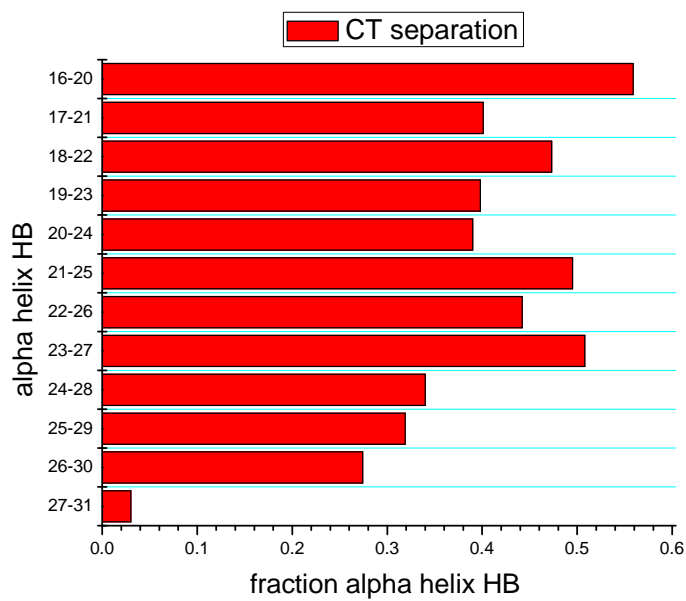

Figure 7. CT separation trajectory (see Table 5). Fractional occurrence over the trajectory of alpha helical hydrogen bonds between the indicated residues for the NT (top panel) and CT (bottom panel) for window 0. The standard errors as a function of block size are shown in the SI Figure S7. Splitting the trajectory into two equal segments and evaluating the ratios of the hydrogen bond fractions shows consistency of the above results (SI Figure S8).

From Figure 7 it is clear that the CT of monomer 1 is as stable as that for the no restraint simulation in Figure 3. The NT shows some slight decrease in the first three alpha helical hydrogen bonds but otherwise the NT is as alpha helical as for the no restraint simulation displayed in Figure 3. The monomer 2 alpha helices are as stable as the no restraint dimer throughout the CT and NT (SI Table S1). Thus, we conclude that separating the CT does not result in melting any part of the monomers. The CT does act as a stable platform for the entire leucine zipper.

**3.5 Restoration of the N terminus dimer.** The simulations summarized in Table 6 were run to attempt to restore the bound dimer structure. They were started from the separated NT simulation end point obtained in Section 3.3.

**Table 6. N terminus restoration simulations [a]**

| Simulation[a] (time) | Purpose | Result | Window T | Window exchange[c] |
|---|---|---|---|---|
| Restore (3X8 ns) [b] | Restore NT to dimer | Dimer restored and some HBs | W_0 1.0 | 0<--->1 0.67 |
| Restore (3X8 ns) [c] | | | W_1 0.990 | 1<--->2 0.93 |
| | | | W_2 0.988 | 2<--->3 0.81 |
| | | | W_3 0.973 | 3<--->4 0.86 |
| | | | W_4 0.968 | 4<--->5 0.79 |
| | | | W_5 0.952 | 5<--->6 0.86 |
| | | | W_6 0.947 | 6<--->7 0.83 |
| | | | W_7 0.931 | |

[a] Half harmonic exterior restraints between Met2-Met2', Leu5-Leu5', Val9-Val9' and Leu12-Leu12'
[b] W_0 FC=0.0; W_I (I=1-7) FC=1.0 kcal/mol/ Å$^2$
[c] W_0 FC=0.0; W_I (I=1-7) FC=0.2 kcal/mol/ Å$^2$

In these simulations, half harmonic exterior restraints, acting between Met2-Met2', Leu5-Leu5', Val9-Val9' and Leu12-Leu12', were applied, except for window 0. Initially, the same set of residues as in the N terminus separation simulations (see Table 4) were tried but they did not lead to the dimer separation decreasing sufficiently. In the first of the simulations summarized in Table 6, the force constants were set to 1.0 kcal/mol/ Å$^2$ with the goal of encouraging the NT parts of the monomers to approach. In this simulation the 0<-->1 exchange probability is poor. However, it did lead to a reduction in the monomer-monomer NT distances. In the second set of simulations, that continue from the end of the first, the force constant was reduced to 0.2 kcal/mol/ Å$^2$ with the goal of keeping the 0<--->1 exchange probability large once the NT residues were already close, so strongly interacting, as shown in Table 6.

In spite of the good exchange probability for window 0 (W_0) it also is important to have the itineration of the configurations into and out of W_0 also be robust. That is, there should be a "random walk" of all the configurations for W_0. The window notation is a proxy for the Hamiltonian parameters used; here, the restraint potential and effective temperature. The configurations ("replicas") track the MD coordinates. Figure 8 displays the itineration of the different replicas into and out of W_0, over a short time interval for clarity. It does show that all the configurations do wander through the W_0 system on a time scale short compared with the total simulation time. Thus, we may conclude that the method is functioning well to accelerate sampling even in the W_0 system, which has a force constant of zero. The other windows that all share the same restraint force constant have similar behavior.

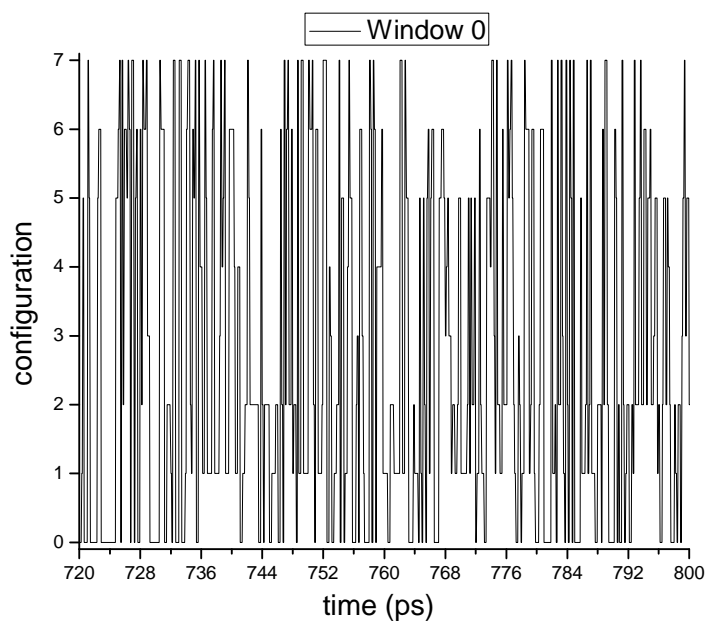

Figure 8. The plot indicates, for a given window (restraint potential), here W_0, which configuration (replica) is visiting W_0 at each instant in time. Only a short time interval is shown for clarity – it is exemplary of the total simulation.

Key to this stage of the simulation is that the force constant for the system labeled as W_0 is zero. This also is the system that corresponds to T=275, the "ambient" temperature. Thus, this system corresponds to the unperturbed, ambient temperature. All the data presented below will be for this unperturbed, T=275 system. The role of the other windows, assuming the exchange probabilities and the itinerations are sufficient, is to provide configurations that are more bound-dimer-like. Note, too, that the restraint potential for the closer-in branch of the restraint potential is zero in all systems, with the desire to encourage those monomer residues to adjust to their mutual presence; that is, to not prejudice the close-in sampling in any system.

A comparison of a snapshot from window 0 of this simulation with the crystal structure is shown in Figure 9. The forces between residues that serve to bind the NT part of the dimer are able to act and, consequently, restore the distances that are found in the crystal structure and in the no restraints simulations of Section 3.2. The restored dimer is stable over the MD trajectory. There is some unraveling of the first few monomer 2 residues but it should be again noted that the first two residues are not stable. Indeed, the nine CA-CA' x-ray compared with the average and standard error distances of this trajectory now are: 6.31 (11.31±0.036), 6.34 (6.92±0.022), 5.47 (5.88±0.013), 6.11 (6.27±0.013), 5.60 (5.51±0.007), 6.20 (6.00±0.005), 5.87 (5.50±0.007), 6.43 (6.07±0.005), and 6.12 (6.2±0.015). Thus the Met2-Met2' dimer interaction is broken but the others are essentially as present as they are in the no restraint simulation discussed in Section 3.3.

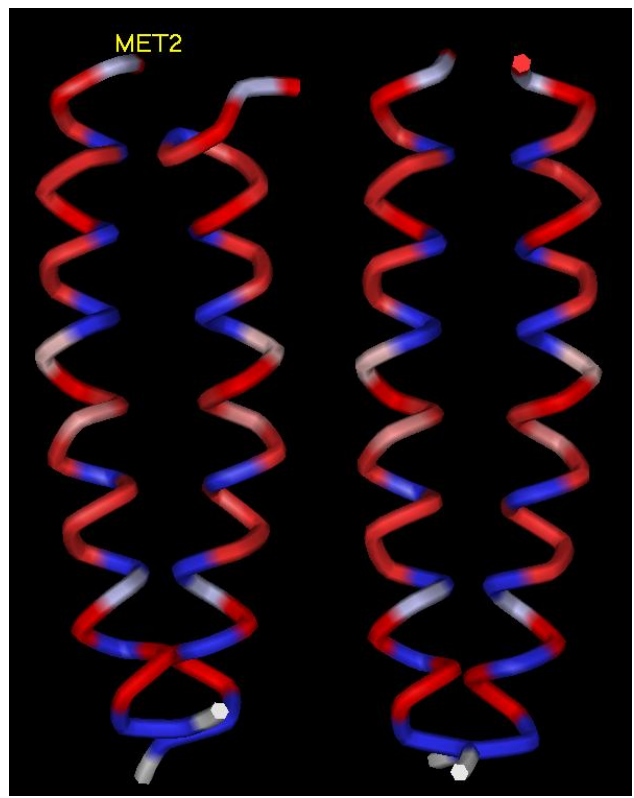

Figure 9. Comparison of the crystal structure (left) with that of a snapshot of the restored dimer structure (right). The color scheme displays hydrophobic residues in blue and hydrophilic in red. The NT structure of monomer 1 is restored but there is some unraveling of the NT of monomer 2 in this snapshot.

However, while the restored structure may appear similar to the crystal structure and the no restraint stable dimer simulations, an examination of the residue 2-8 alpha helical hydrogen bonds, displayed in Figure 10, show that the restored dimer structure has a low percentage of these NT hydrogen bonds, in contrast with the roughly 35-50% hydrogen bonding in the no restraint simulation of the dimer shown in Figure 3. There are actually a number of i-i+3 and i-i+2 hydrogen bonds found, though at low population, between Met2-Leu5, Lys3-Leu5, Leu5-Lys8, and Lys3-Asp7. The seven other NT 1-4 hydrogen bonds between residues 5-9 to 11-15 have similar fractions to those found for the no restraint simulation, with the exception that the Leu5-Val9 is somewhat less present, indicating some residual lack of alpha helical interaction.

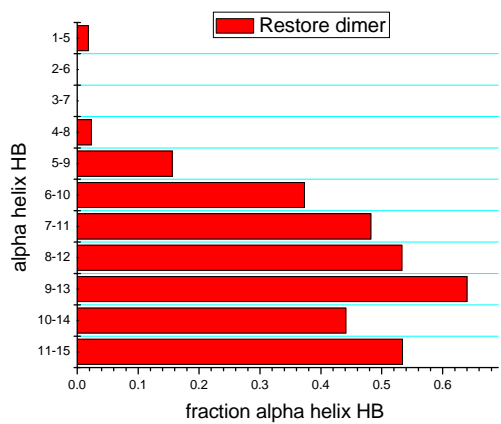

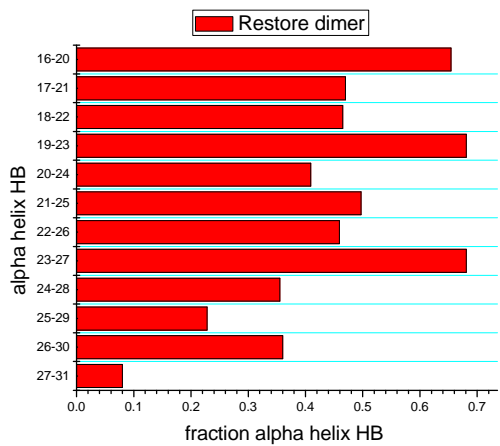

Figure 10. Fractional occurrence over the restore dimer trajectory of alpha helical hydrogen bonds between the indicated residues for the NT (top panel) and CT (bottom panel) for window 0. The restored dimer structure has a low percentage of NT Met2-Leu6 through Gln4-Lys8 alpha helical hydrogen bonds. This is in contrast with the roughly 25-50% hydrogen bonding for this range in the no restraint simulation of the dimer shown in Figure 3. The standard errors as a function of block size are shown in the SI Figure S9. Splitting the trajectory into two equal segments and evaluating the ratios of the hydrogen bond fractions shows consistency of the above results (SI Figure S10).

Why is alpha helical hydrogen bonding spanning residues 2-6 to 4-8 not restored here, relative to the no restraint simulations discussed in Section 3.2? A possible explanation centers on the presence of the many ionized residues in this part of the NT (see the sequence given in the Introduction). A schematic of these residues is shown in Figure 11. For the indicated charged residues, there is extensive side-chain side-chain interaction possible within and between the monomers. These interactions can either compete with or promote the re-formation of the alpha-helical hydrogen bonding for this part of the NT. Note that the first two residues, Arg1 and Met2 are unstructured in all the simulations, and are not present in the crystal structure or the structure from which we initiated the simulations.

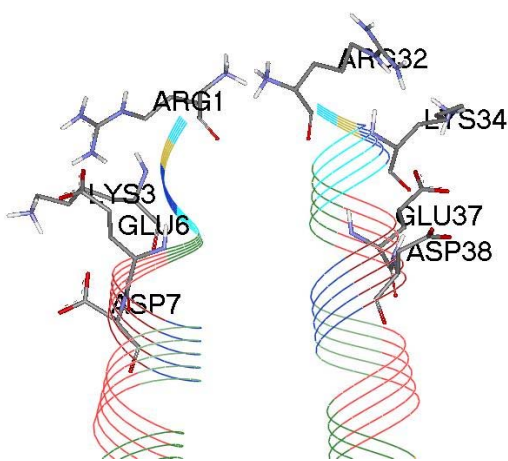

Figure 11. Potential intra- and inter-monomer salt bridging residues in the beginning part of the N terminal sequence.

To investigate this possibility, the fractions of intra- and inter-monomer salt bridges among these charged residues in the no restraint (Section 3.2) and restoration of the N terminus of the dimer simulations (Section 3.5) are displayed in Figure 12. It seems from the general decrease in the number of significant salt bridges in going from the no restraint to the restoration simulation that this diminution of salt bridging may impede the re-formation of some monomer alpha-helical hydrogen bonds.

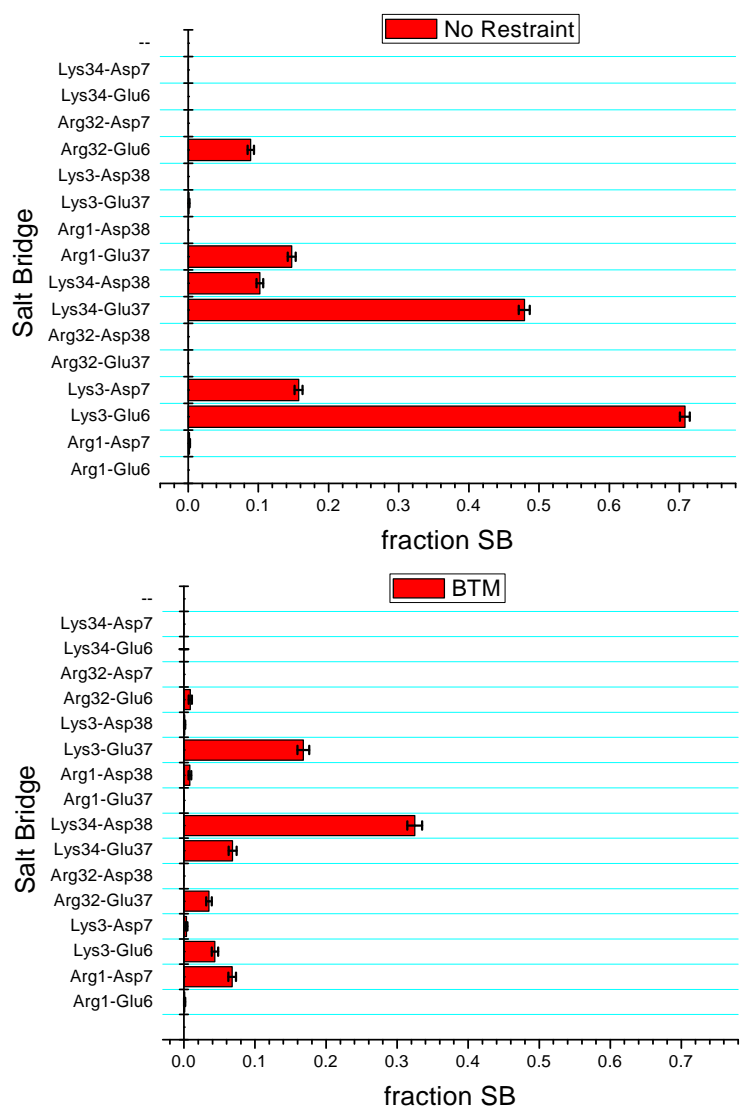

Figure 12 Comparison of monomer-monomer salt bridges between the no restraint (Section 3.2) and the restoration of the N terminus dimer (Section 3.5) simulations. Percentages over the trajectories are given along with the standard errors obtained from block averaging. For the restoration of the N terminus dimer (Section 3.5) the data for window 0 corresponds to ambient T and no constraints. (See Table 6).

## 4. DISCUSSION AND CONCLUSIONS

In this work, a combination of HTREM that provides enhanced sampling by use of effectively higher temperatures than ambient and HTREM_RMS that, in addition, adds strategically designed restraints to encourage sampling of desired regions of configuration space was introduced. HTREM accelerates sampling in a more efficient manner than temperature REM because fewer degrees of freedom are scaled. The design of the added restraints is quite flexible. Here, half harmonic restraints were used to encourage both separation and bringing together of parts of the leucine zipper dimer. For separations, the exterior part of the potential that induces no forces permits freer exploration of separated regions and the finite-force branch induces more sampling of the separated regions. For bringing together, the acting exterior forces serve to keep the close-in region in contact more than in the absence of these forces. Naturally, such restraints can be used in other contexts where the desire is to enhance sampling of certain regions of configuration space without prejudicing the exploration in the desired region. Note that conventional harmonic restraints can over determine the sampling in this sense. It is also convenient to use a restraint that is a sum over restraint potentials, to again allow for freer sampling, in contrast with a set of specific restraints.

An extensive simulation (total of 64 ns) of the dimer using HTREM with a large range of effective temperatures showed that the dimer is quite stable. The intra-monomer distances are well maintained and the alpha helical hydrogen bonding of each monomer is not disrupted. Monitoring these hydrogen bonds showed that the NT ones, between residues 3-7 to 11-15 and the CT ones between residues 12-16 to 26-30 are fractionally present between ~25-65% of the trajectory, for the ambient temperature. Similar results are obtained for monomer 2 (SI Table S1). What fractional percentage is obtained of course depends on the definition of a hydrogen bond, here heavy atom distance less than 3.0 Å and OHN angle between 0 and 30°. There is no distinct difference between these percentages for the beginning of the NT – hydrogen bonds between residues 2-6 to 4-8 and the remaining ones in the NT or the CT. In this sense, once bound, the monomers are "uniformly" stable helices, with the first two NT and the last CT residues undergoing greater fluctuations than other residues.

Separating the NT monomers was accomplished with half harmonic restraints, serving to push them apart progressively with the largest distance targeted at the end of the NT. That strategy was designed to mimic unzipping the dimer. The force constant used was sufficient to accomplish this separation, as shown in Figure 4. Most importantly, monitoring the alpha helical hydrogen bonds between residues 2-6 to 4-8 shows that they are absent and the 5-9 to 7-11 are much reduced, relative to the equilibrium no restraint simulation. Thus, this part of the NT has lost its helical character. In contrast, applying an analogous separating scheme to the CT, did not lead to melting of either the CT or the NT. For the CT separation, a more drastic procedure was used than for the NT separation, as summarized in Table 5. Namely, first a large range of effective temperatures was used to accelerate the sampling and after this a smaller temperature range was used for data collection. Note that the window exchanges are excellent for all these simulations, indicating that the HTREM method was effective in enhancing sampling. The combination of NT monomer melting upon NT separation and stability of the dimer upon CT separation supports the trigger sequence hypothesis.[15] It should be noted that MD force fields are approximate and that the differences found between these two simulations, while potentially accidental, does provide strong evidence for the hypothesis.[39-41]

In the restoration of the dimer simulation, half harmonic restraints with the forces on the exterior range were used. First with a relatively large force constant to encourage the NT monomers to approach within an interaction distance, and then with a small force constant to minimize the disturbances from the applied forces. In both these simulation sets, no force constant is applied to window 0; thus, the sampling for this window corresponds to the unperturbed, ambient temperature. In the smaller force constant simulation, all exchange probabilities are high and, in addition, the itineration of configurations into and out of window 0 is robust. Thus, the benefit from the HTREM_MS sampling should be maximized. The result of the simulation is to bring the monomers back to a similar geometric structure as in the starting dimerized leucine zipper, as shown in Figure 9. However, the alpha helical hydrogen bonding is not restored. As discussed in Section 3.5, this may be due to the loss of some intra- and inter-monomer salt bridges among the charged residues involved that are present in the no restraints dimer simulation.

**Supporting Information**: Table S1 summarizes the monomer 2 alpha helical hydrogen bonding fractions of all the simulations for the N and C terminal domains. Figure S1 has hydrogen bond fractions of the no restraint trajectory with a less strict hydrogen bond definition. Figures S2, S4, S7 and S9 have block averaged standard errors for all the simulations. Figures S3, S5, S8 and S10 compare averages over the first and second halves of each simulation trajectory. Figure S6 has the CT separation trajectory hydrogen bond fractions for a simulation at the smaller effective temperature range.

**Acknowledgements**: The author acknowledges the High Performance Computer Center at Michigan State University for computational cycles.

**References**

1. Burkhard, P.; Stetefeld, J.; Strelkov, S. V., Coiled Coils: A Highly Versatile Protein Folding Motif. *Trends Cell Biol.* **2001,** *11*, 82-88.

2. Gruber, M.; Lupas, A. N., Historical Review: Another 50th Anniversary - New Periodicities in Coiled Coils. *Trends Biochem. Sci.* **2003,** *28*, 679-685.

3. Lai, J. R.; Fisk, J. D.; Weisblum, B.; Gellman, S. H., Hydrophobic Core Repacking in a Coiled-Coil Dimer Via Phage Display: Insights into Plasticity and Specificity at a Protein-Protein Interface. *J. Am. Chem. Soc.* **2004,** *126*, 10514-10515.

4. Lupas, A. N.; Gruber, M., The Structure of Alpha-Helical Coiled Coils. In *Adv. Protein. Chem.*, 2005; Vol. 70, pp 37–38.

5. Mason, J. M.; Arndt, K. M., Coiled Coil Domains: Stability, Specificity, and Biological Implications. *Chembiochem* **2004,** *5*, 170-176.

6. Mason, J. M.; Hagemann, U. B.; Arndt, K. M., Role of Hydrophobic and Electrostatic Interactions in Coiled Coil Stability and Specificity. *Biochemistry* **2009,** *48*, 10380-10388.

7. Moutevelis, E.; Woolfson, D. N., A Periodic Table of Coiled-Coil Protein Structures. *J. Mol. Biol.* **2009,** *385*, 726-732.

8. Oakley, M. T.; Garibaldi, J. M.; Hirst, J. D., Lattice Models of Peptide Aggregation: Evaluation of Conformational Search Algorithms. *J. Comput. Chem.* **2005,** *26*, 1638-1646.

9. Woolfson, D. N., The Design of Coiled-Coil Structures and Assemblies. In *Adv. Protein. Chem.*, 2005; Vol. 70, pp 80–106.

10. Ellenberger, T., Getting a Grip on DNA Recognition: Structures of the Basic Region Leucine Zipper, and the Basic Region Helix-Loop-Helix DNA-Binding Domains. *Curr. Opin. Struct. Biol.* **2004,** *4*, 12-21.

11. Crick, F. H. C., The Packing of Alpha-Helices - Simple Coiled-Coils. *Acta Crystallogr.* **1953,** *6*, 689-697.

12. Ibarra-Molero, B.; Zitzewitz, J. A.; Matthews, C. R., Salt-Bridges Can Stabilize but Do Not Accelerate the Folding of the Homodimeric Coiled-Coil Peptide GCN4-P1. *J. Mol. Biol.* **2004,** *336*, 989-996.

13. Marti, D. N.; Bosshard, H. R., Electrostatic Interactions in Leucine Zippers: Thermodynamic Analysis of the Contributions of Glu and His Residues and the Effect of Mutating Salt Bridges. *J. Mol. Biol.* **2003,** *330*, 621-637.

14. Matousek, W. M.; Ciani, B.; Fitch, C. A.; Garcia-Moreno, B.; Kammerer, R. A.; Alexandrescu, A. T., Electrostatic Contributions to the Stability of the GCN4 Leucine Zipper Structure. *J. Mol. Biol.* **2007,** *374*, 206-219.

15. Steinmetz, M. O.; Jelesarov, I.; Matousek, W. M.; Honnappa, S.; Jahnke, W.; Missimer, J. H.; Frank, S.; Alexandrescu, A. T.; Kammerer, R. A., Molecular Basis of Coiled-Coil Formation. *Proc. Natl. Acad. Sci. U. S. A.* **2007,** *104*, 7062-7067.

16. Bunagan, M. R.; Cristian, L.; DeGrado, W. F.; Gai, F., Truncation of a Cross-Linked GCN4-P1 Coiled Coil Leads to Ultrafast Folding. *Biochemistry* **2006,** *45*, 10981-10986.

17. Dragan, A. I.; Privalov, P. L., Unfolding of a Leucine Zipper Is Not a Simple Two-State Transition. *J. Mol. Biol.* **2002,** *321*, 891-908.

18. Meisner, W. K.; Sosnick, T. R., Barrier-Limited, Microsecond Folding of a Stable Protein Measured with Hydrogen Exchange: Implications for Downhill Folding. *Proc. Natl. Acad. Sci. U. S. A.* **2004,** *101*, 15639-15644.

19. Meisner, W. K.; Sosnick, T. R., Fast Folding of a Helical Protein Initiated by the Collision of Unstructured Chains. *Proc. Natl. Acad. Sci. U. S. A.* **2004,** *101*, 13478-13482.

20. Nikolaev, Y.; Pervushin, K., NMR Spin State Exchange Spectroscopy Reveals Equilibrium of Two Distinct Conformations of Leucine Zipper GCN4 in Solution. *J. Am. Chem. Soc.* **2007,** *129*, 6461-6469.

21. Liu, Y.; Chagagain, P. P.; Parra, J. L.; Gerstman, B. S., Lattice Model Simulation of Interchain Protein Interactions and the Folding Dynamics and Dimerization of the GCN4 Leucine Zipper. *J. Chem. Phys.* **2008,** *128*.

22. Mohanty, D.; Kolinski, A.; Skolnick, J., De Novo Simulations of the Folding Thermodynamics of the GCN4 Leucine Zipper. *Biophys. J.* **1999,** *77*, 54-69.

23. Vieth, M.; Kolinski, A.; Brooks, C. L.; Skolnick, J., Prediction of the Folding Pathways and Structure of the GCN4 Leucine-Zipper. *J. Mol. Biol.* **1994,** *237*, 361-367.

24.	Vinals, J.; Kolinski, A.; Skolnick, J., Numerical Study of the Entropy Loss of Dimerization and the Folding Thermodynamics of the GCN4 Leucine Zipper. *Biophys. J.* **2002,** *83*, 2801-2811.

25.	Rojas, A. V.; Liwo, A.; Scheraga, H. A., Molecular Dynamics with the United-Residue Force Field: Ab Initio Folding Simulations of Multichain Proteins. *J. Phys. Chem. B* **2007,** *111*, 293-309.

26.	Choi, Y. H.; Yang, C. H.; Kim, H. W.; Jung, S., Leucine Zipper as a Fine Tuner for the DNA Binding; Revisited with Molecular Dynamics Simulation of the Fos-Jun Bzip Complex. *Bull. Korean Chem. Soc.* **1999,** *20*, 1319-1322.

27.	Yadav, M. K.; Leman, L. J.; Price, D. J.; Brooks, C. L.; Stout, C. D.; Ghadiri, M. R., Coiled Coils at the Edge of Configurational Heterogeneity. Structural Analyses of Parallel and Antiparallel Homotetrameric Coiled Coils Reveal Configurational Sensitivity to a Single Solvent-Exposed Amino Acid Substitution. *Biochemistry* **2006,** *45*, 4463-4473.

28.	Gorfe, A. A.; Ferrara, P.; Caflisch, A.; Marti, D. N.; Bosshard, H. R.; Jelesarov, I., Calculation of Protein Ionization Equilibria with Conformational Sampling: Pk(a) of a Model Leucine Zipper, GCN4 and Barnase. *Proteins-Struct. Func. Gen.* **2002,** *46*, 41-60.

29.	Pinero, A.; Villa, A.; Vagt, T.; Koksch, B.; Mark, A. E., A Molecular Dynamics Study of the Formation, Stability, and Oligomerization State of Two Designed Coiled Coils: Possibilities and Limitations. *Biophys. J.* **2005,** *89*, 3701-3713.

30.	Missimer, J. H.; Steinmetz, M. O.; Jahnke, W.; Winkler, F. K.; van Gunsteren, W. F.; Daura, X., Molecular-Dynamics Simulations of C- and N-Terminal Peptide Derivatives of GCN4-P1 in Aqueous Solution. *Chemistry & Biodiversity* **2005,** *2*, 1086-1104.

31.	Ubbink, M., The Courtship of Proteins: Understanding the Encounter Complex. *FEBS Lett.* **2009,** *583*, 1060-1066.

32.	Keskin, Z.; Gursoy, A.; Ma, B.; Nussinov, R., Principles of Protein-Protein Interactions: What Are the Preferred Ways for Proteins to Interact? *Chem. Rev.* **2008,** *108*, 1225-1244.

33.	Dyson, H. J.; Wright, P. E., Intrinsically Unstructured Proteins and Their Functions. *Nat Rev Mol Cell Bio* **2005,** *6*, 197-208.

34.	Habchi, J.; Tompa, P.; Longhi, S.; Uversky, V. N., Introducing Protein Intrinsic Disorder. *Chem. Rev.* **2014,** *114*, 6561-6588.

35. O'Shea, E. K.; Klemm, J. D.; Kim, P. S.; Alber, T., X-Ray Structure of the GCN4 Leucine Zipper, a 2-Stranded, Parallel Coiled Coil. *Science* **1991,** *254*, 539-544.

36. Brandon, C.; Tooze, J., *Introduction to Protein Structure*. second ed.; Garland Publishing: New York, 1999.

37. Kammerer, R. A.; Schulthess, T.; Landwehr, R.; Lustig, A.; Engel, J.; Aebi, U.; Steinmetz, M. O., An Autonomous Folding Unit Mediates the Assembly of Two-Stranded Coiled Coils. *Proc. Natl. Acad. Sci. U. S. A.* **1998,** *95*, 13419-13424.

38. Lee, D. L.; Lavigne, P.; Hodges, R. S., Are Trigger Sequences Essential in the Folding of Two-Stranded A-Helical Coiled-Coils? *J. Mol. Biol.* **2001,** *306*, 539-553.

39. Bhandari, Y. R.; Chapagain, P. P.; Gerstman, B. S., Lattice Model Simulations of the Effects of the Position of a Peptide Trigger Segment on Helix Folding and Dimerization. *J. Chem. Phys.* **2012,** *137*, 105103.

40. Chapagain, P. P.; Liu, Y. X.; Gerstman, B. S., The Trigger Sequence in the GCN4 Leucine Zipper: Alpha-Helical Propensity and Multistate Dynamics of Folding and Dimerization. *J. Chem. Phys.* **2008,** *129*, 175103.

41. Zitzewitz, J. A.; Ibarra-Molero, B.; Fishel, D. R.; Terry, K. L.; Matthews, C. R., Preformed Secondary Structure Drives the Association Reaction of GCN4-P1, a Model Coiled-Coil System. *J. Mol. Biol.* **2000,** *296*, 1105-1116.

42. Su, L.; Cukier, R. I., Hamiltonian and Distance Replica Exchange Method Studies of Met-Enkephalin. *J. Phys. Chem. B* **2007,** *111*, 12310-12321.

43. Su, L.; Cukier, R. I., Hamiltonian Replica Exchange Method Studies of a Leucine Zipper Dimer. *J. Phys. Chem. B* **2009,** *113*, 9595-9605.

44. Cukier, R. I., A Hamiltonian Replica Exchange Method for Building Protein-Protein Interfaces Applied to a Leucine Zipper. *J. Chem. Phys.* **2011,** *134*, 045104.

45. Garcia, A. E.; Sanbonmatsu, K. Y., Exploring the Energy Landscape of a Beta Hairpin in Explicit Solvent. *Proteins-Structure Function and Genetics* **2001,** *42*, 345-354.

46. Geyer, C. J., Markov Chain Monte Carlo Maximum Likelihood. In *Computing Science and Statistics: Proceedings of the 23rd Symposium on the Interface* Keramidas, E. M., Ed. Interface Foundation: Fairfax Station, 1991.

47. Hansmann, U. H. E.; Okamoto, Y., Numerical Comparisons of Three Recently Proposed Algorithms in the Protein Folding Problem. *J. Comput. Chem.* **1997,** *18*, 920-933.

48. Sugita, Y.; Okamoto, Y., Replica-Exchange Molecular Dynamics Method for Protein Folding. *Chem. Phys. Lett.* **1999,** *314*, 141-151.

49. Dickson, A.; Ahlstrom, L. S.; Brooks, C. L., Coupled Folding and Binding with 2d Window-Exchange Umbrella Sampling. *J. Comput. Chem.* **2016,** *37*, 587-594.

50. Park, S.; Kim, T.; Im, W., Transmembrane Helix Assembly by Window Exchange Umbrella Sampling. *Phys. Rev. Letts.* **2012,** *108*.

51. Lou, H. F.; Cukier, R. I., Molecular Dynamics of Apo-Adenylate Kinase: A Principal Component Analysis. *J. Phys. Chem. B* **2006,** *110*, 12796-12808.

52. van Gunsteren, W. F.; Billeter, S. R.; Eising, A. A.; Hünenberger, P. H.; Krüger, P.; Mark, A. E.; Scott, W. R. P., *Biomolecular Simulation: The Gromos96 Manual and User Guide*. Vdf hochschulverlag AG an der ETH: Zürich, 1996.

53. Lou, H.; Cukier, R. I. *Analyzer*, 2.0; East Lansing, 2008.

54. Flyvbjerg, H.; Petersen, H. G., Error-Estimates on Averages of Correlated Data. *J. Chem. Phys.* **1989,** *91*, 461-466.

# Supplementary Information

A Trigger Sequence in a Leucine Zipper Aids its Dimerization; Simulation Results

Robert I. Cukier

**Table S1.**
**Monomer 2 NT alpha helical hydrogen bonds**

| HB (M1 notation) | HB | No restraint | NT separation | CT separation | Restore dimer |
|---|---|---|---|---|---|
| 1-5 | 32-36 | 0.284 | 0.109 | 0.247 | 0.000 |
| 2-6 | 33-37 | 0.457 | 0.371 | 0.559 | 0.253 |
| 3-7 | 34-38 | 0.570 | 0.376 | 0.660 | 0.168 |
| 4-8 | 35-39 | 0.497 | 0.215 | 0.477 | 0.211 |
| 5-9 | 36-40 | 0.496 | 0.235 | 0.554 | 0.471 |
| 6-10 | 37-41 | 0.327 | 0.254 | 0.436 | 0.335 |
| 7-11 | 38-42 | 0.587 | 0.422 | 0.516 | 0.519 |
| 8-12 | 39-43 | 0.627 | 0.457 | 0.483 | 0.615 |
| 9-13 | 40-44 | 0.675 | 0.447 | 0.660 | 0.686 |
| 10-14 | 41-45 | 0.352 | 0.508 | 0.564 | 0.335 |
| 11-15 | 42-46 | 0.488 | 0.500 | 0.350 | 0.513 |

**Monomer 2 CT alpha helical hydrogen bonds**

| HB (M1 notation) | HB | No restraint | NT separation | CT separation | Restore dimer |
|---|---|---|---|---|---|
| 16-20 | 47-51 | 0.684 | 0.222 | 0.467 | 0.687 |
| 17-21 | 48-52 | 0.393 | 0.210 | 0.415 | 0.389 |
| 18-22 | 49-53 | 0.496 | 0.421 | 0.576 | 0.464 |
| 19-23 | 50-54 | 0.677 | 0.484 | 0.397 | 0.697 |
| 20-24 | 51-55 | 0.411 | 0.332 | 0.422 | 0.387 |
| 21-25 | 52-56 | 0.560 | 0.559 | 0.566 | 0.494 |
| 22-26 | 53-57 | 0.451 | 0.506 | 0.421 | 0.522 |
| 23-27 | 54-58 | 0.673 | 0.603 | 0.538 | 0.668 |
| 24-28 | 55-59 | 0.352 | 0.328 | 0.375 | 0.330 |
| 25-29 | 56-60 | 0.233 | 0.342 | 0.354 | 0.225 |
| 26-30 | 57-61 | 0.381 | 0.390 | 0.376 | 0.371 |
| 27-31 | 58-62 | 0.006 | 0.342 | 0.033 | 0.039 |

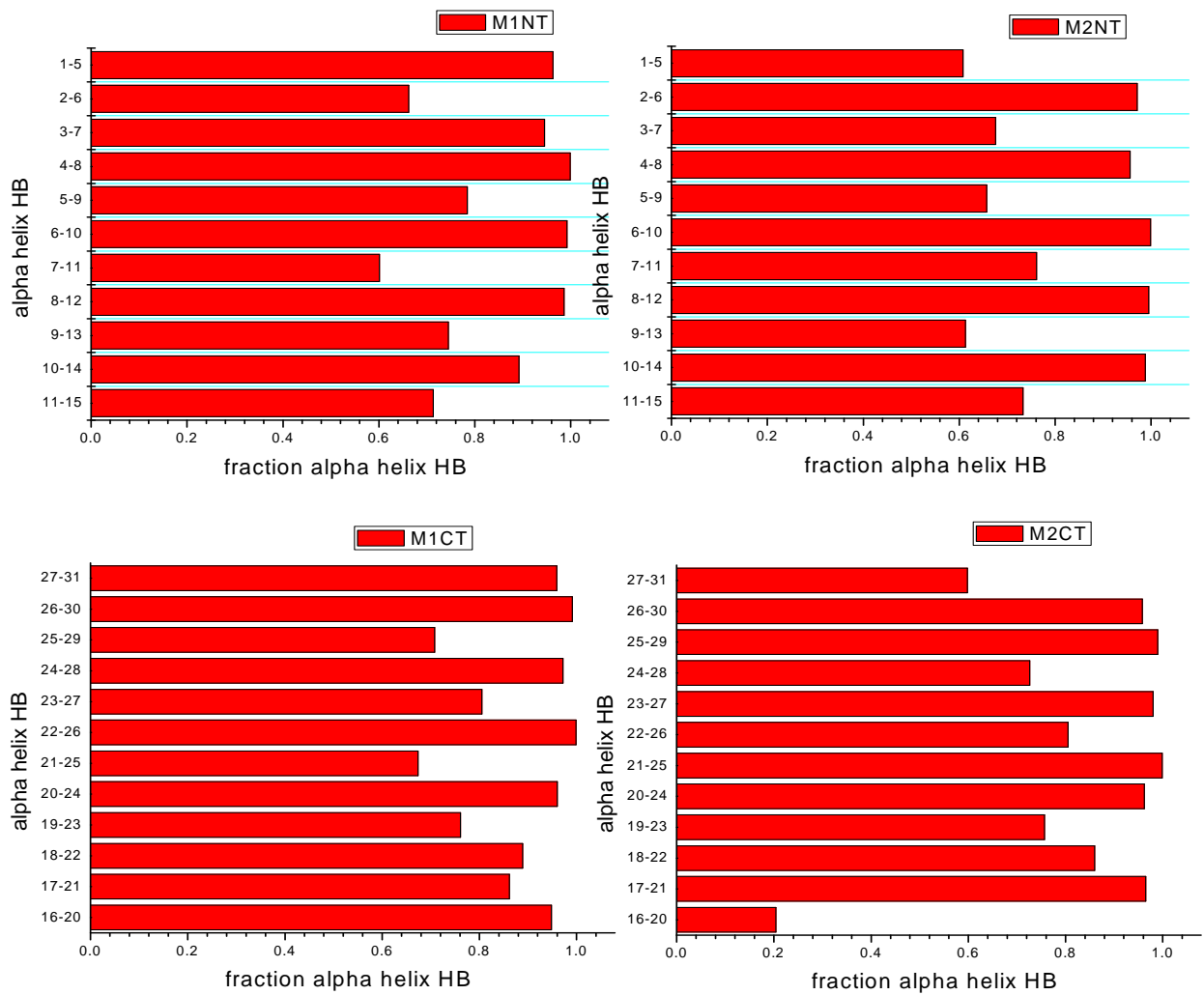

Figure S1. Fractional occurrence over the no restraint dimer trajectory (Table 3) of alpha helical hydrogen bonds between the indicated residues for the NT monomer 1 and 2 (top panel) and CT monomer 1 and 2 (bottom panel). The alpha helical hydrogen bonds between the nitrogen of the residue *n*+4 N-H and the carbonyl of the residue *n* C-O are defined by backbone nitrogen-to-carbonyl carbon distance less than 3.5 Å and corresponding OHN angle between 0 and 30°.

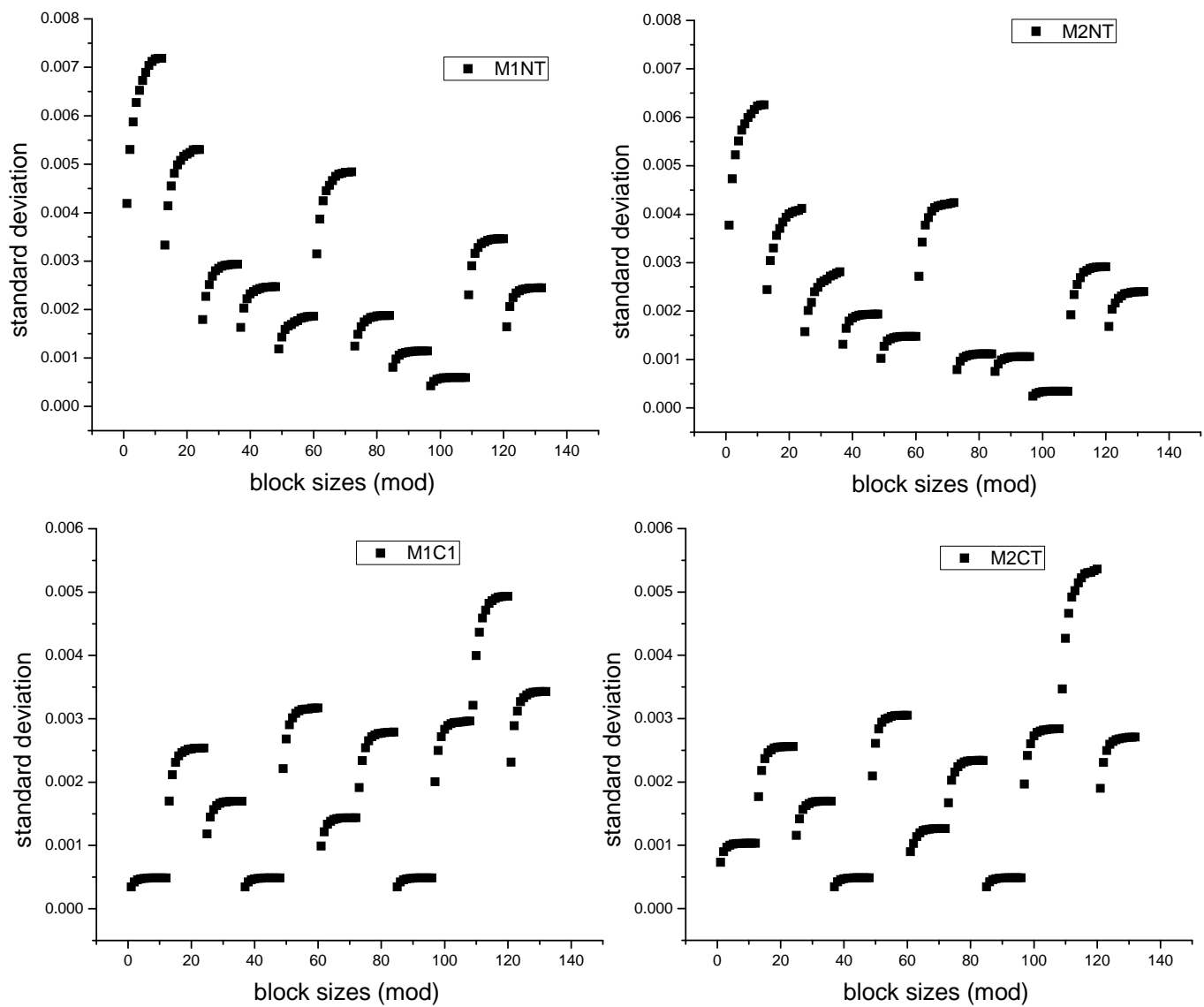

Figure S2. Block averaged standard errors for the presence versus absence during the trajectory of all hydrogen bonds whose average percentage presence are displayed in Figure 3. The block sizes (mod 11) show the standard errors of blocks of size $2^1$ to $2^{11}$ trajectory samples, for the hydrogen bonds in each subdomain: the M1 NT (monomer 1 N terminal) M1 CT (monomer 1 C terminal) hydrogen bonds as defined in Figure 3. The monomer 2 hydrogen bonds are defined analogously. Plateau values are typically reached for block sizes corresponding to less than 100 ps, indicating convergence of the hydrogen bond data on this time scale. The scale of the standard errors are all very small compared with their average values.

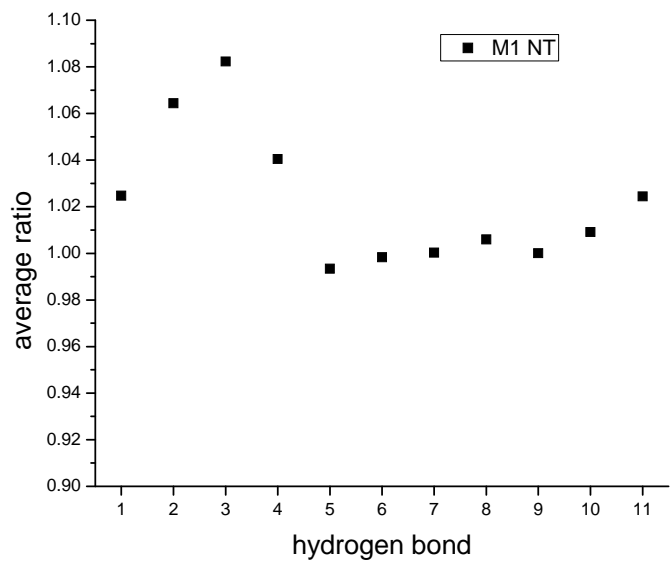

Figure S3. Ratio of the 11 NT hydrogen bond fraction averages of the first and second half of the trajectory used to construct Figure 3 for monomer 1. The ratio is essentially unity indicating that the trajectory is equilibrated.

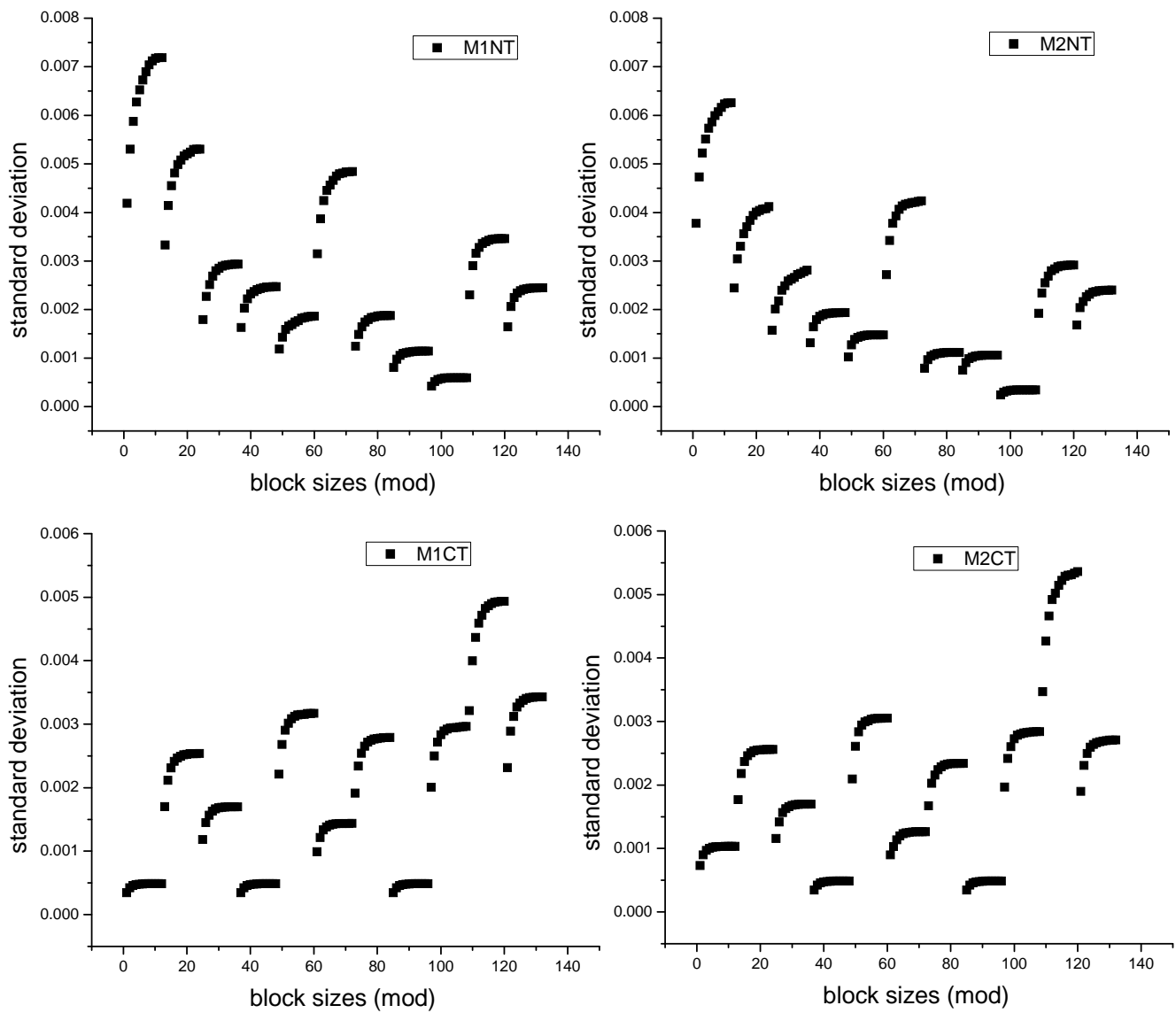

Figure S4. Block averaged standard errors for the presence versus absence during the trajectory of all hydrogen bonds whose average percentage presence are displayed in Figure 5. The block sizes (mod 11) show the standard errors of blocks of size $2^1$ to $2^{11}$ trajectory samples for the hydrogen bonds in each subdomain: the M1 NT (monomer 1 N terminal) M1 CT (monomer 1 C terminal) hydrogen bonds as defined in Figure 3. The monomer 2 hydrogen bonds are defined analogously. Plateau values are typically reached for block sizes corresponding to less than 100 ps, indicating convergence of the hydrogen bond data on this time scale. The scale of the standard errors are all very small compared with their average values.

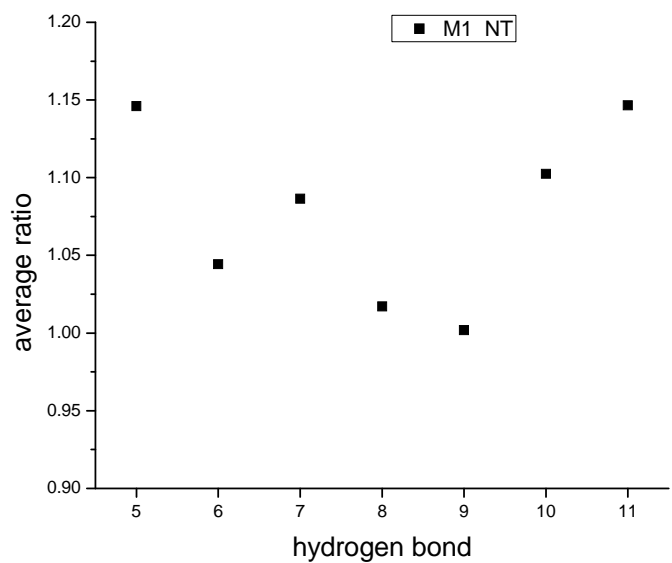

Figure S5. Ratio of the 6 finite percentage NT hydrogen bond fraction averages of the first and second half of the trajectory used to construct Figure 5 for monomer 1. The ratio is essentially unity indicating that the trajectory is equilibrated.

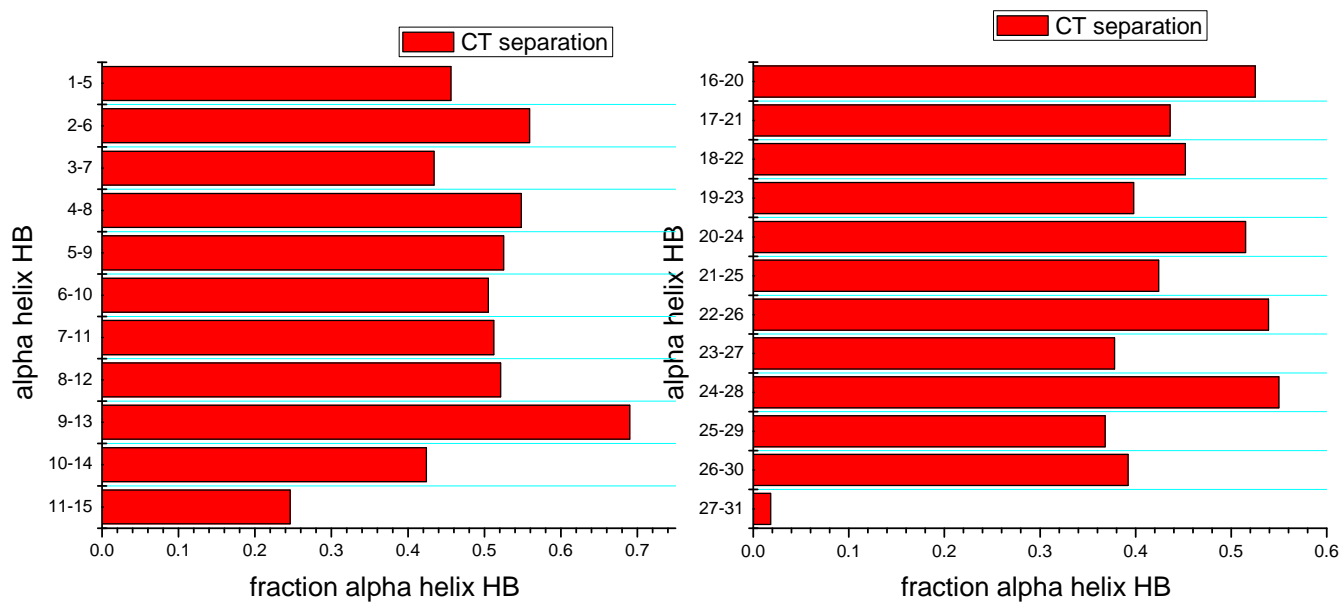

Figure S6. CT separation trajectory based on a 2 X 8 ns run at the small T range (the second set of Table 5 window T's). Fractional occurrence over the trajectory of alpha helical hydrogen bonds between the indicated residues for the NT (top panel) and CT (bottom panel). They are quite similar with the exception of somewhat more stable hydrogen bonds for the first two NT hydrogen bonds (1-5 and 2-6).

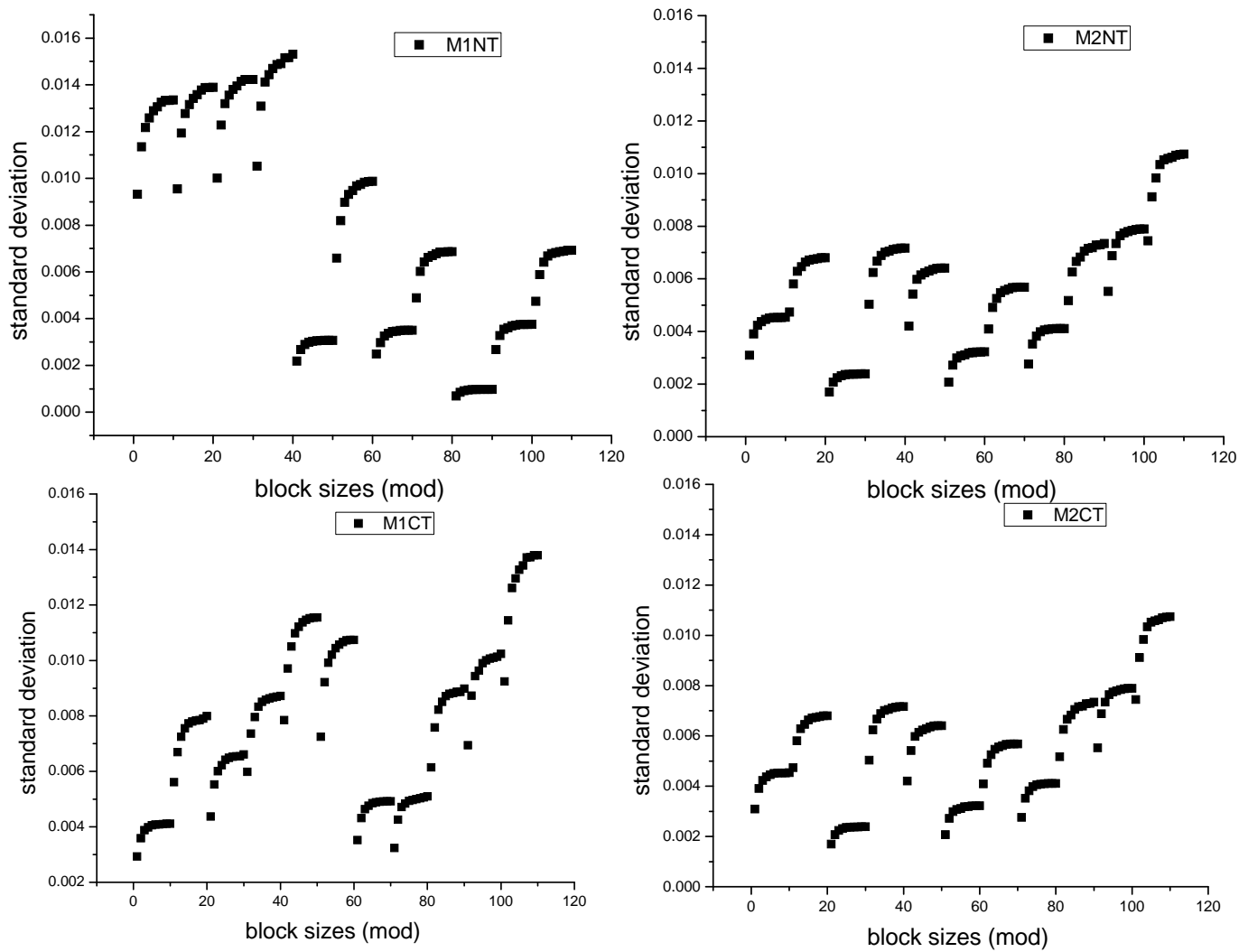

Figure S7. Block averaged standard errors for the presence versus absence during the trajectory of all hydrogen bonds whose average percentage presence are displayed in Figure 7. The block sizes (mod 10) show the standard error of blocks of size $2^1$ to $2^{10}$ trajectory samples, for the hydrogen bonds in each subdomain: the M1 NT (monomer 1 N terminal) M1 CT (monomer 1 C terminal) hydrogen bonds as defined in Figure 3. The monomer 2 hydrogen bonds are defined analogously. Plateau values are typically reached for block sizes corresponding to less than 100 ps, indicating convergence of the hydrogen bond data on this time scale. The scale of the standard errors are all very small compared with their average values.

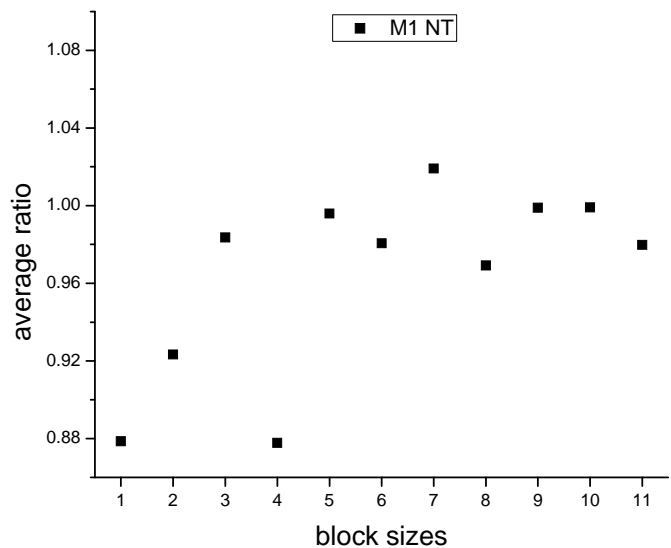

Figure S8. Ratio of the 11 NT hydrogen bond fraction averages of the first and second half of the trajectory used to construct Figure 7 for monomer 1. The first three HBs are low population. The ratio is essentially unity with some deviation for the fourth HB, indicating that the trajectory is equilibrated.

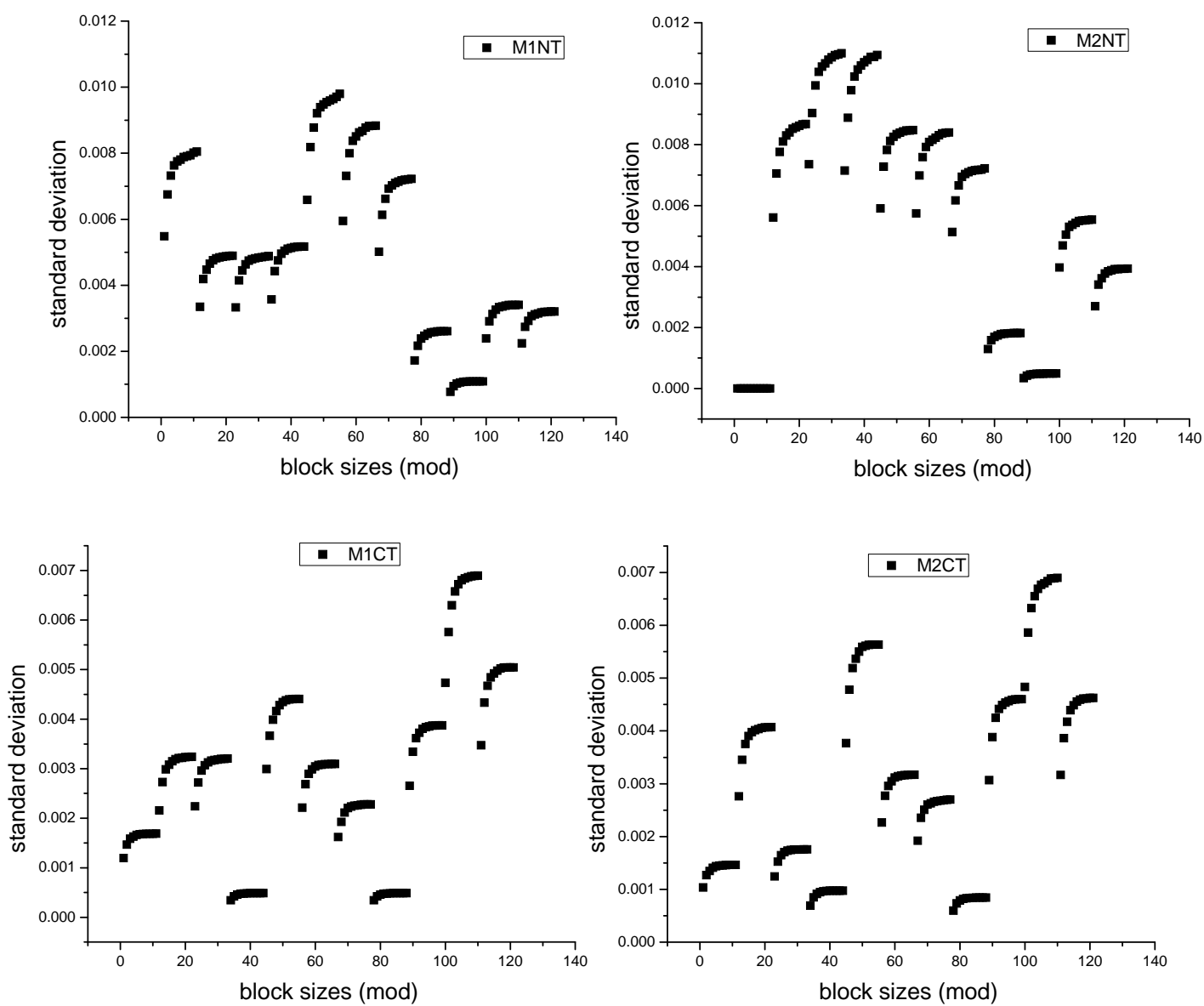

Figure S9. Block averaged standard errors for the presence versus absence during the trajectory of all hydrogen bonds whose average percentage presence are displayed in Figure 10. The block sizes (mod 11) show the standard error of blocks of size $2^1$ to $2^{11}$ trajectory samples, for the hydrogen bonds in each subdomain: the M1 NT (monomer 1 N terminal) M1 CT (monomer 1 C terminal) hydrogen bonds as defined in Figure 3. The monomer 2 hydrogen bonds are defined analogously. Plateau values are typically reached for block sizes corresponding to less than 100 ps, indicating convergence of the hydrogen bond data on this time scale. The scale of the standard error are all very small compared with their average values.

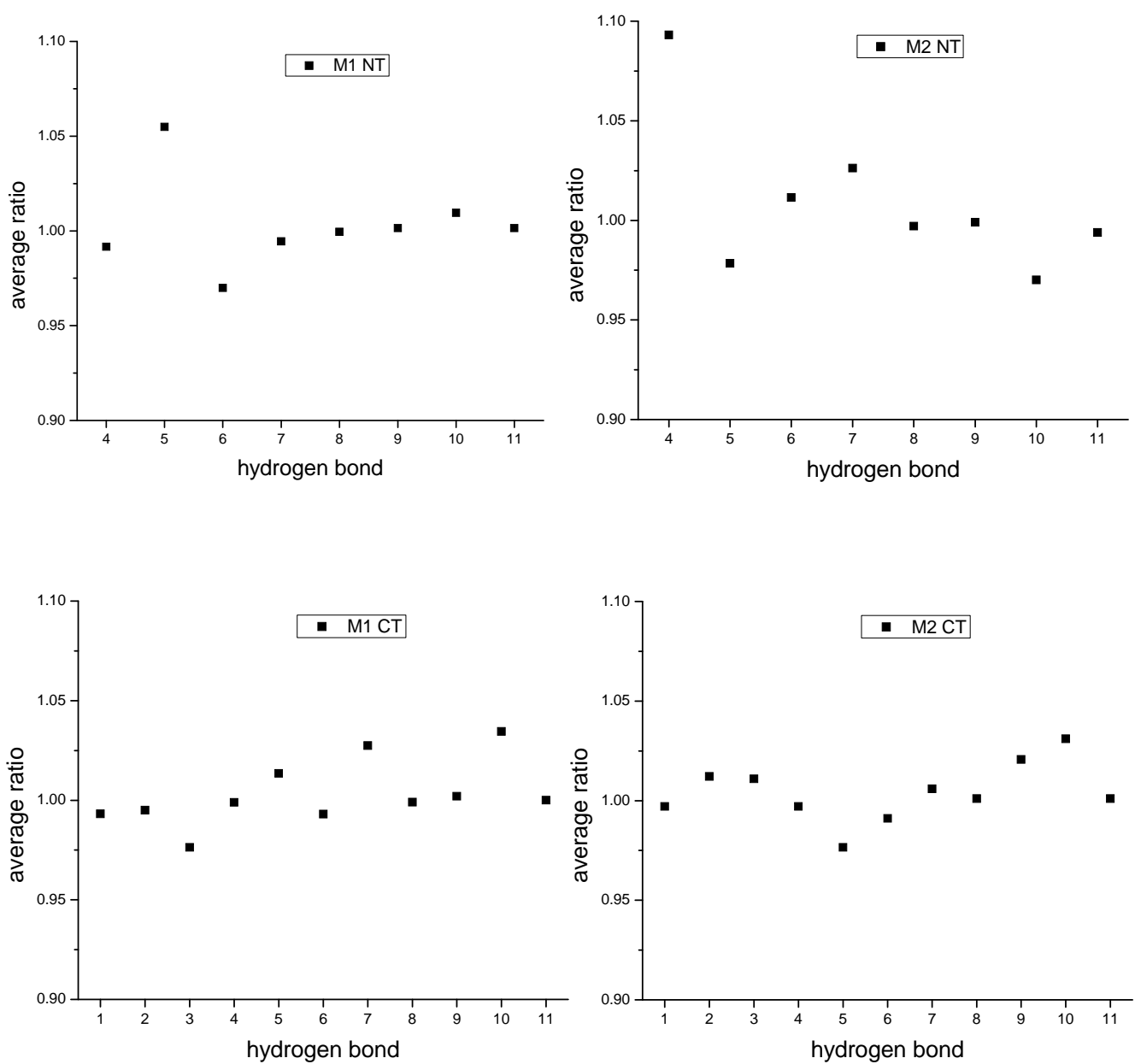

Figure S10. Ratio of the averages of the first and second half of the trajectory used to construct Figure 10 for the monomer 1 and 2 NT hydrogen bond percentages. The ratio is essentially unity (the first three HBs are low percentage), indicating that the trajectory is equilibrated. For the CT the hydrogen bonds are all strong except for the last one that is hardly present (see Figure 10).